\documentclass[useAMS,usenatbib]{mn2e}

\usepackage{graphicx}
\usepackage{subfigure}
\title[]{VLT-FORS2 optical imaging and spectroscopy of 9 luminous type 2 AGN at
  0.3$<$z$<$0.6: I. Ionized gas nebulae\thanks{Based on observations carried out at the European Southern Observatory
(Paranal, Chile) with FORS2 on VLT-UT1 (programme 087.B-0034)}}
\author[Humphrey et al.]{A. Humphrey$^{1}$,
  M. Villar-Mart\'{i}n$^{2,3}$, C. Ramos Almeida$^{4}$,
  C.~N. Tadhunter$^{5}$, \newauthor
S. Arribas$^{2,3}$, P.~S. Bessiere$^{6}$, A. Cabrera-Lavers$^{4}$\\
$^{1}$Instituto de Astrof\'{i}sica e Ci\^encias do Espa\c{c}o, Universidade do Porto, CAUP, Rua das Estrelas, PT4150-762 Porto, Portugal\\
$^{2}$Centro de Astrobiolog\'{i}a (INTA-CSIC), Carretera de Ajalvir, km 4, 28850 Torrej\'on de Ardoz, Madrid, Spain\\
$^3$Astro-UAM, UAM, Unidad Asociada CSIC, Facultad de Ciencias, Campus de Cantoblanco, E-28049, Madrid, Spain \\
$^{4}$Instituto de Astrof\'{i}sica de Canarias, (IAC) V\'{i}a L\'actea s/n, E-38205, La Laguna, Tenerife, Spain \\
$^{5}$Department of Physics and Astronomy, University of Sheffield, Sheffield S3 7RH, UK \\
$^{6}$Universidad de Concepci\'on, Departamento de Astronom\'{i}a, Casilla 160-C, Concepci\'on, Chile
}

\begin{document}

\date{Accepted 2015 September 17.  Received 2015 September 17; in original form 2015 July 30.}

\pagerange{\pageref{firstpage}--\pageref{lastpage}}
\pubyear{2013}

\maketitle

\label{firstpage}

\begin{abstract}
We  present  optical imaging and long slit spectroscopic observations
of 9 luminous type 2 AGNs within the redshift range 0.3$<$z$<$0.6
based on VLT-FORS2 data. Most objects (6/9) are high luminosity
Seyfert 2, and three are type 2 quasars (QSO2), with our sample
extending to lower luminosity than previous works. 

Seven out of nine objects (78\%) show morphological evidence for
interactions or mergers in the form of disturbed morphologies and/or
peculiar features such as tidal tails, amorphous halos, or compact
emission line knots. The detection rate of morphological evidence for
interaction is consistent with those found during previous studies of
QSO2 at similar $z$, suggesting that the merger rate is independent of
AGN power at the high end of the AGN luminosity function. 

We find the emission line flux spatial profiles are often dominated
by the often spatially unresolved central source. In addition, all
but one of our sample is associated with much fainter, extended line
emission. We find these extended emission line structures have a
variety of origins and ionization mechanisms: star forming companions,
tidal features, or extended ionized nebulae. AGN related processes
dominate the excitation of the nuclear gas. Stellar photoionization
sometimes plays a role in extended structures often related to
mergers/interactions.

\end{abstract}

\begin{keywords}

\end{keywords}

\section{Introduction}
The coevolution and interplay of galaxies with their nuclear black
hole is a topic of key importance for understanding galaxy
evolution. Negative feedback from an active galactic nucleus (AGN hereinafter)
may impact on the evolution of the stellar and gaseous components of
the host galaxy, by heating or expelling cold gas that may otherwise
have formed stars or fed the AGN, potentially hindering mass
assembly activity. Feedback such as this has been invoked to explain
the deviation of the observed galaxy stellar mass function from
the theoretical function (White \& Frenk 1991; Puchwein \& Springel
2013), to account the observed correlation 
between the black hole mass and the stellar mass of the stellar
spheroidal component of the host galaxy (Magorrian et
al. 1998), and to facilitate the transformation of dusty, obscured
galaxies to unobscured, optical galaxies (e.g. Sanders et al. 1988;
Bessiere et al. 2014). Although some negative feedback 
clearly does take place in galaxies as they undergo a
phase of AGN activity, the dominant mechanism and strength of this
feedback remains a matter of debate in the literature. Positive
feedback, where star formation is triggered or enhanced due to
AGN-related processes, has also been suggested to take place under
some circumstances (Rees 1989; Silk 2013), with a few tentative detections in the 
literature (Croft et al. 2006; Stroe et al. 2014).

Triggering is another key issue for understanding AGN activity and its
relationship with the evolution of the host galaxy. Gas-rich major
mergers are thought able to trigger powerful nuclear activity, insofar
as they provide a mechanism to displace large quantities of cold gas
into the central few kiloparsecs of one (or more) of the merging
galaxies (e.g. Heckman et al. 1986), and there is a growing body of
evidence to support this idea (e.g. Tadhunter et al. 2011; Ramos
Almeida et al. 2011; 2012; Bessiere et al. 2012; 2014). 

\begin{table*}
\centering
\caption{The sample of type 2 quasars (QSO2,
  $l_{O3}$=log($\frac{L_{[OIII]}}{L_{\odot}})>8.3$ ) and high
  luminosity Seyfert 2 (HSy2, $l_{O3}<$8.3).} 
\begin{tabular}{lllllllll}
\hline
Target & Short name & z & log($\frac{L_{[OIII]}}{L_{\odot}})$ & SDSS &FIRST $S_{1.4GHz}$ & NVSS $S_{1.4GHz}$ & Class \\
& &   & &  (g mag)  & (mJy) & (mJy) \\
\hline
SDSS J090307.83+021152.2 & SDSS J0903+02 &  0.329 & 8.79 &  19.5 & 22.5$\pm$0.1 & 24.2$\pm$0.8 & QSO2\\
SDSS J092318.06+010144.8 & SDSS J0923+01 &  0.386 & 8.78 & 20.4 & 1.0$\pm$0.2 & 2.3$\pm$0.5 &QSO2 \\
SDSS J095044.69+011127.2 & SDSS J0950+01 & 0.404 & 8.22 &  21.1 & 1.7$\pm$0.1 & &  HSy2  \\
SDSS J101403.49+024416.4 & SDSS J1014+02 &  0.573 & 8.29 &  22.6 & 2.4$\pm$0.1 & & HSy2 \\
SDSS J101718.63+033108.2 & SDSS J1017+03 &   0.453 & 8.27 & 21.4 & 2.8$\pm$0.2 & &HSy2 \\
SDSS J124749.79+015212.6 & SDSS J1247+01 & 0.427 & 8.23 & 21.2 & 8.0$\pm$0.1 & 34.1$\pm$1.6 &HSy2\\
SDSS J133633.65-003936.4 & SDSS J1336-00 & 0.416 & 8.64 & 20.7 & $\le$0.5 & &QSO2 \\
SDSS J141611.77-023117.1 & SDSS J1416-02 &  0.305 & 8.03 & 21.2 & 1.4$\pm$0.1 & &HSy2 \\
SDSS J145201.73+005040.2 & SDSS J1452+00 &  0.315 & 7.82 & 21.3 & 0.5$\pm$0.2 &  &HSy2\\
\hline
\end{tabular}
\label{tab:sample}
\end{table*}

\begin{table*}
\centering
\caption{Summary of the VLT  FORS2 observations.  Columns: (1) source
  name; (2) source redshift; (3) date of the observation (run in April 2011); (4) type of
observation, where NB indicates narrow band imaging, IB indicates
intermediate band imaging, BB indicates broad band imaging, LSS
indicates long slit spectroscopy, and HST indicates HST WFPC2 imaging;
(5) filter, or grism and slit combination; (6) 
the position angle of the long slit, anti-clockwise from north (north
through east); (7) the exposure time on source
of the observation; (8) the FWHM of the seeing disc as measured from stars in the broad band images; (9) the average FWHM seeing conditions 
calculated over the exposure time as measured by the Differential Image Motion Monitor (DIMM) station; (10)  the
kiloparsec to arcsec conversion.} 
\begin{tabular}{lllllllllll}
\hline
Name & z  & Night & Obs. & Filter / Grism & PA ($^{\circ}$) &
Exp. (s) & Seeing (\arcsec)$^{im}$ & Seeing(\arcsec)$^{dimm}$  & kpc/\arcsec \\
(1) & (2) & (3) & (4) & (5) & (6) & (7) & (8) & (9) & (10) \\
\hline
SDSS J0903+02 & 0.329 & 28 & BB & v\_HIGH &   & 730 & 0.81$\pm$0.03 & 0.88$\pm$0.05 &  4.71 \\
 & & 28 & NB & H\_Alpha/4500+61&   & 900 & 0.95$\pm$0.02 & 0.89$\pm$0.05\\
 & & 28 & LSS & 600RI 1.3 \arcsec~slit & -65.5 (PA1)& 4200 & & 1.00$\pm$0.29 \\
 & & 28 & LSS & 600RI 1.3\arcsec~slit & 63.4 (PA2)& 1400 & & 1.05$\pm$0.22 \\
& & 28 & LSS & 300I 1.0 \arcsec~slit & -65.5 (PA1)& 2800 & &  0.93$\pm$0.24 \\
& &      & HST & F814W & & 1200 & & \\
\hline
SDSS J0923+01 & 0.386 & 28 & BB & v\_HIGH &   & 600 & 0.66$\pm$0.06 & 0.79$\pm$0.10 & 
5.23 \\
& & 28 & LSS & 300I 1.0\arcsec~slit & 40.9 & 3578 &   &  0.85$\pm$0.15\\
& &      & HST & F814W & & 1200 & & \\
\hline
SDSS J0950+01 & 0.404 & 27 & BB & v\_HIGH &   & 600 & 0.96$\pm$0.05 & 1.09$\pm$0.12 & 5.38 \\
& & 27 & IB & FILT\_530\_25 &   & 900 & 1.05$\pm$0.04 & 1.16$\pm$0.11 \\
& & 27 & LSS & 600RI 1.3\arcsec~slit& 9.9 & 4200 & &   1.05$\pm$0.13\\
\hline
SDSS J1014+02 & 0.573 & 26 & BB & v\_HIGH &   & 600 & 0.81$\pm$0.04 &   0.78$\pm$0.07  &
6.52 \\
& & 26 & NB & HeI+53 &   & 900 & 0.64$\pm$0.03 & 0.89$\pm$0.08 \\
& & 26 & LSS & 600RI 1.0\arcsec~slit & -5.9 (PA1) & 2142 &  & 0.58$\pm$0.07\\
& & 26 & LSS & 600RI 1.0\arcsec~slit & 42.1 (PA2) & 1400 &  & 0.76$\pm$0.10 \\
\hline
SDSS J1017+03 & 0.453 & 28 & BB & v\_HIGH &   & 600 & 0.74$\pm$0.03 & 0.69$\pm$0.02 &
5.76 \\
& & 28 & LSS & 300I 1.0\arcsec~slit & 37.7 & 2100 & & 0.71$\pm$0.06\\
\hline
SDSS J1247+01  & 0.427 & 25 & BB & v\_HIGH &   & 600 & 1.20$\pm$0.02 &  1.35$\pm$0.08    & 5.56 \\
& & 25 & NB & HeI+53 &   & 900 & 1.24$\pm$0.05 & 1.36$\pm$0.12 \\
& & 25 & LSS & 600RI 1.3\arcsec~slit & 60.7 & 1443 & & 1.50$\pm$0.21 \\ 
\hline
SDSS J1336-00 & 0.416 &   26 & BB & v\_HIGH &   & 600 & 0.66$\pm$0.05 & 0.69$\pm$0.07 & 5.47 \\
& & 26 & IB & FILT\_530\_25 &   & 300 &  0.63$\pm$0.02 & 0.84$\pm$0.07 \\
& & 26 & LSS & 600RI 1.0\arcsec~slit & 57.4 & 2800 & & 0.62$\pm$0.07\\
\hline

SDSS J1416-02 & 0.305 & 27 & BB & v\_HIGH &   & 600 & 1.10$\pm$0.05 & 1.01$\pm$0.07 & 4.47 \\
& & 27 & IB & FILT\_485\_37 &   & 900 & 1.04$\pm$0.04 & 0.95$\pm$0.04  \\
& & 27 & LS & 600RI 1.3\arcsec~slit & 90.0 & 2800 & & 0.84$\pm$0.08 \\
\hline
SDSS J1452+00 & 0.315 & 28 & BB & v\_HIGH &  & 600 & 0.60$\pm$0.04  & 0.61$\pm$0.03 & 
4.56 \\ 
& & 28 & LSS & 300I 1.0\arcsec~slit & -64.6 & 2100 &  & 0.59$\pm$0.05\\
\hline
\end{tabular}
\label{tab:obs}
\end{table*}

Radio quiet, type 2 quasars 
are a potential goldmine of information to improve our understanding
of several important aspects of the evolution of massive galaxies,
including the triggering and impact of AGN activity therein.
One of their main advantages is that the fortuitous 
obscuration by an optically thick structure of the highly luminous
central engine, which can otherwise outshine the entire stellar
content of the host galaxy, affords a cleaner picture of the galaxy.  
Moreover, compared to their radio-loud cousins (powerful radio galaxies),
the space density of radio-quiet type 2 quasars is roughly an order of
magnitude higher (Reyes et al. 2008), making them far more
representative of powerful 
active galaxies.  In addition, the absence of powerful radio jets
removes substantial 
ambiguity about whether or not observed properties of the host galaxy
are induced by the radio jets, which often impacts studies of radio
loud quasars and radio galaxies.  However, their lack of powerful
radio emission or highly luminous 
optical emission meant that, even after their existence was
hypothesized during efforts to unify the seemingly disparate varieties
of active galaxy radio quiet, type 2 quasars remained relatively
elusive for decades (see e.g. Halpern et al. 1999).

It was not until the advent of the Sloan Digital 
Sky Survey (SDSS) that type 2 radio quiet quasars were identified in
significant numbers.  Using selection criteria designed to find
galaxies that contain gas photoionized by a powerful but obscured AGN,
Zakamska et al. (2003) were able to identify $\sim$300 candidate type
2 quasars within the range 0.3$<$z$<$0.8, of which 85\% are
radio-quiet (see also Reyes et al. 2008).  Subsequent observations at
various wavelengths confirmed that these galaxies are indeed type 2
quasars, with unobscured luminosities that places them among the most
luminous quasars at similar redshifts (Zakamska et al. 2004, 2006;
Ptak et al. 2006; Lal \& Ho 2010). Complementary to the above, radio
quiet type 2 quasars have now also been found in significant numbers
by using data from infrared, X-ray and radio surveys to select heavily
obscured AGN (e.g., Ohta et al. 1996; Norman et al. 2002;
Mart\'{i}nez-Sansigre et al. 2005; 2006a; 2006b), or by using
redshifted ultraviolet emission 
lines (e.g. Ly$\alpha$) to select type 2 quasars at z$\ga$2
(Alexandroff et al. 2013). At the time of writing, almost 1000
optically-selected type 2 quasars have now been identified (e.g. Reyes
et al. 2008). 

The bright optical emission lines that facilitate the identification
of type 2 quasars at low to intermediate redshift also carry
information about the properties of the gaseous component of the host
galaxy. Spatially extended ($\ga$10 kpc) regions of warm
(T$\sim$10,000 K) ionized gas are frequently detected (e.g. Humphrey
et al. 2010; Villar-Mart\'{i}n et al. 2011a, hereafter VM11a; Liu et
al. 2013a,b; Harrison et al. 2014; McElroy et al. 2015). In most cases, the
excitation of this gas is dominated by 
photoionization by the hard radiation field of the central AGN, with
an occasional contribution from young stars (Villar-Mart\'{i}n et
al. 2008) or, in at least one case, a contribution from shock
ionization (Humphrey et al. 2010). Many of these quasars show evidence 
for outflows in their nuclear narrow line region (Villar-Mart\'{i}n et
al. 2011b (VM11b hereafter), 2012; Humphrey et al. 2010; Liu et al. 2013b; see also
e.g. Arribas et al. 2014). In addition, 
some intermediate redshift type 2 quasars also contain a large mass of 
molecular gas, as detected in CO line emission (Villar-Mart\'{i}n et
al. 2013a,b; Rodr\'{i}guez et al. 2014).

The host galaxies of the
majority of intermediate redshift type 2 
quasars show elliptical morphologies, with around half showing
morphological disturbances identifiable with galaxy merger or
interaction (Villar-Mart\'{i}n et al. 2012; Bessiere et
al. 2012). Although their stellar populations are dominated by old
stars, young or post-starburst populations are also prevalent
(e.g. Bian 2007; Bessiere et al. 2016, in prep.), suggesting roughly 
synchronous triggering of the AGN and starburst activity during a
merger event, albeit with rather different delays from the starburst
to AGN triggering from quasar to quasar (see, e.g., Bessiere et
al. 2014). Arguments from environmental clustering amplitudes suggest 
that all massive elliptical galaxies go through a short-lived phase
as a radio-quiet quasar (Ramos Almeida et al. 2013). 

In 2008 we started an observational program with the Very Large Telescope (VLT)
based on optical imaging and spectroscopy of QSO2, with the goal of
investigating, quantifying and characterizing 1) the existence of
ionized outflows, 2) the incidence of interactions/mergers and 3) the
existence, properties and origin of extended ionized structures in the most powerful radio
quiet type 2 active galaxies. In this paper, we extend the work
published in four earlier papers (Villar Mart\'\i n et al. 2010; 2012;
VM11a; VM11b), focusing now on a somewhat less luminous sample.
Here we concentrate on the optical morphology and merger status of the
quasar host galaxies, and the spatial distribution and excitation of
narrow-line emitting gas therein. In a future paper (Villar-Mart\'{i}n
et al., in prep.; Paper II) we will present a detailed kinematic
analysis and discuss the properties of ionized gas outflows. 

The paper is organized as follows. The sample is described in
Sect. 2. The observations and data reduction are explained in Sect. 3.   
Analysis and results are presented in  Sect. 4 both for the individual
objects. The overall results are discussed in Sect. 5 and the
conclusions are presented in Sect. 6.

We adopt $H_{0}$=71 km s$^{-1}$ Mpc$^{-1}$, $\Omega_{\Lambda}$=0.73 and
$\Omega_{m}$=0.27.  At the redshifts of our sample, this gives an
arcsec to kpc conversion ranging from 4.47 kpc/arcsec to 6.52 kpc/arcsec.

\begin{figure*}
\includegraphics{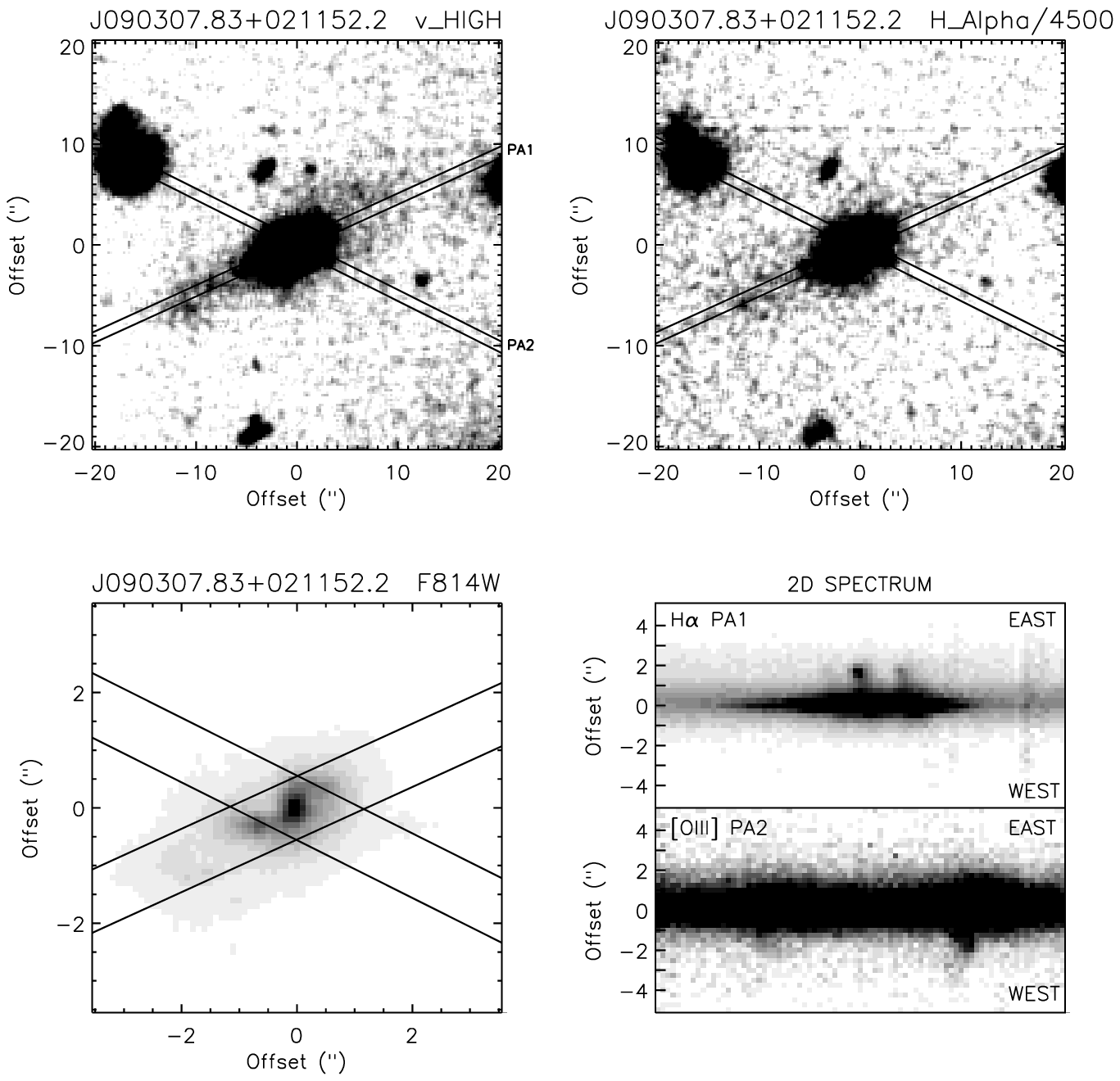}
\vspace{5.25in}
\caption{Images, the [OIII] $\lambda\lambda$4959,5007 line profile,
  and sections of the two-dimensional spectrum of SDSS
  J090307.83+021152.2. North is up and East is left in this and all remaining figures.} 
\label{fig_0903}
\end{figure*}

\section{The Sample}

The sample consists of 9 luminous type 2 AGN   selected from the SDSS sample of high luminosity type 2 AGN selected by \cite{rey08}.  They are objects  with
narrow ($<$2000 km s$^{-1}$) emission lines without underlying broad components for the recombination lines suggestive of a BLR, and with line ratios characteristic of non-stellar ionizing radiation.
 Some basic information is presented in Table  \,\ref{tab:sample}.
 All objects have redshift z$\sim$0.3-0.6 such that there was an
adequate narrow- or intermediate-band VLT-FORS2 filter containing one
of the strongest emission lines in the optical spectrum ([O II]$\lambda$3727
or [O III]$\lambda$5007). The [OIII] luminosities are in the range $l_{O3}$=log$\frac{L_{[OIII]}}{L_{\odot}}$=7.8-8.8 (luminosities taken from \cite{rey08} Vizier online catalogue). We selected our targets to have large emission line equivalent widths.

For completeness, in Table   \,\ref{tab:sample} we also give the flux
density of each target at 1.4 GHz, as measured by the Faint Images of
the Radio Sky at Twenty-Centimeters survey (FIRST: Becker et al. 1995)
and, where available, as measured by the NRAO VLA Sky Survey (NVSS:
Condon et al. 1998). Issues such as the radio-loudness of the targets
will be explored in Paper II. 

The sample is similar to that studied by VM11a and VM11b. The main difference is that here we 
focus on somewhat less luminous objects. Our previous sample had a range of [OIII] luminosities with median value corresponding to  $l_{O3}$=8.84, while the current sample has 8.27.
\cite{zak03} and \cite{rey08} used slightly different criteria to select SDSS QSO2 based
on their spectroscopic properties, imposing $l_{O3}>$8.5 and $>$8.3 respectively.
Although these numbers are only indicative,  and there might  be a gradual transition of properties with increasing luminosity,
we have divided our sample in QSO2 and  high luminosity Seyfert 2s (HSy2, hereafter) adopting \cite{rey08} criteria (i.e.  a transition value of $l_{O3}$=8.3). The classification does not change
if we use criteria \cite{zak03} instead.   As a result, while the
sample of VM11a and VM11b consisted
of 12 QSO2 and 2 HSy2, the current sample contains 3 QSO2 and 6 HSy2.

\section{Observations and data reduction}

\subsection{Very Large Telescope observations}

The VLT observations were made during the dark nights of 25-28 April 2011 at
the Antu unit of the Very Large Telescope, as part of the programme 087.B-0034. The Focal Reducer and
Low Dispersion Spectrograph (FORS2: Appenzeller et al. \citeyear{app98}) was used in
long-slit spectroscopy and imaging modes. The seeing FWHM  varied between $\sim$0.6 and $\sim$1.1\arcsec~ during
 the run (Table  \,\ref{tab:obs}).

For each quasar the observing procedure involved first obtaining an
image through the v\_HIGH broad band filter, and in most cases 
an image through an intermediate or narrow band filter chosen to
contain the redshifted [OII] $\lambda$3727 or [OIII] $\lambda$5007
lines.  Aside from the valuable
scientific information they 
provide, the images were used during the observing run to identify
extended structures and possible companion objects, the
presence and positions of which motivated our choice of slit
position angle during the subsequent long-slit spectroscopic
exposures.  

The spectroscopic observations used one of two  grisms.  
The 600RI grism gave a useful wavelength range of $\sim$5000-8000 \AA,
and a spectral resolution of 5.2$\pm$0.2 \AA~ (FWHM) when using the 1.0\arcsec~
slit, or 7.2$\pm$0.2 \AA~ when using the 1.3\arcsec~ slit. The 300I grism provided
a useful range of $\sim$6000-10000 \AA, and a resolution of 9.6$\pm$0.3 \AA~ with
the 1.0\arcsec~ slit.  The spatial pixel scale is 0.25 arcsec pixel$^{-1}$.The data were processed and calibrated using
 IRAF to apply standard data reduction techniques (see
Villar-Mart\'{i}n et al. \citeyear{vm12} for details).     

A total of ten luminous type 2 AGN at 0.3$<$z$<$0.6 were observed during 
this VLT programme; the results for one of these (SDSS
J143027.66-005614.8) have been published in a separate paper (Villar-Mart\'{i}n
et al. \citeyear{vm12}).  

\subsubsection{{\bf Seeing and slit effects}}
One of the main objectives of this observational programme is to examine the spatial
properties of the narrow emission line gas of our sample (this paper),
and to determine whether extended ionized outflows are present (Paper
II). For this, a careful characterization of the seeing size (FWHM) and shape and its uncertainties is crucial.

The seeing was very variable during the VLT observations (FWHM in the range $\sim$0.6\arcsec-1.3\arcsec).    To account for the uncertainties,
we quote 
in Table  \,\ref{tab:obs} the seeing size measured from the broad band and narrow/intermediate band images using several stars in the field  (column 8) and the average FWHM seeing conditions 
calculated over the exposure time as measured by the Differential Image Motion Monitor (DIMM) station (column 9). The dispersion in these values 
will be carefully taken into account, especially when it could have an impact on our conclusions regarding the spatial extension of a given object. 

In addition, we have reconstructed the spatial profile of the seeing
disk along the slit, using a non-saturated star in images taken
immediately before or after the spectroscopic observation of the
science target. Because our study of the spatial extension of the
ionized gas will be based primarily on [OIII]$\lambda$5007, whenever
possible the narrow or intermediate band image containing this line
was used, in order to minimize the effects of the wavelength
dependence of the seeing profile. In the absence of either a narrow
band or intermediate band image, the broad band image was instead
used. In all cases, the selected star was sufficiently bright to
detect and trace adequately the faint wings of the seeing profile. The
stellar flux was extracted from apertures centered on the stellar
centroid and mimicking the slit: 4 (1.0$\arcsec$) or 5  pixels
(1.3$\arcsec$) wide depending on the slit used for the science
target's spectroscopic observation. Finally, the sky background was
removed from the stellar spatial profiles. 

For some of our science targets, the [OII]$\lambda$3727 line was within the observed spectral
range, and should provide some useful information to complement that
to be obtained using [OIII], provided the wavelength dependence of the
seeing profile is understood and taken into account. To this end, we
have used standard star spectra, obtained during the observing run
(albeit at different air masses), and studied the star's seeing point spread
function as a function of wavelength. We find that the FWHM of the seeing profile
can vary by $\sim$9-16\% between the observed wavelengths of the 
[OII] and [OIII] lines. In our science analysis, this will be taken
into account when relevant. 

The slit width  was chosen between 1.0\arcsec~ and 1.3\arcsec~ 
to reach a compromise between optimizing  the observing time,  obtaining an adequate spectral resolution  and avoiding  significant flux loses. The slit  was often
wider than the seeing disk as a consequence (see Table  \,\ref{tab:obs}). This introduces additional uncertainties for spatially unresolved sources
 on  the kinematic measurements. On one hand, if the objects are clumpy, the image of a spatially unresolved clump will be smaller
 than the slit width and the instrumental profile at that particular spatial position will be narrower
 than the profile measured using the sky or arc lines. This would lead to an underestimation of the intrinsic FWHM of the 
emission lines, which is the result of subtracting the instrumental profile in quadrature. The resulting uncertainties have  been carefully taken into account
and will be mentioned when relevant.

\subsection{Auxiliary $HST$ data}
SDSS J0903+02 and  SDSS J0923+01  have also been
observed using the Wide Field and Planetary Camera 2 on the Hubble Space
Telescope ($HST$).  These data were retrieved from the Hubble Legacy Archive (HLA). The images were obtained
for the HST program with identification 10880 and principal investigator Henrique Schmitt. In both 
cases the broad band F814W filter was used.

\section{Analysis and Results}

Figs. ~\ref{fig_0903}-\ref{fig_1452} show our
FORS2 images along with the corresponding $HST$ image where
available. Sections of the two-dimensional spectra are also shown to
highlight extended nebular features, if detected. 

In addition, in Fig. ~\ref{spat} we show emission line spatial
profiles (usually [OIII] $\lambda$5007), with the spatial point spread
function during each observation also plotted for comparison. The
spatial profiles of the emission lines were derived by extracting
a one-dimensional spectrum at each spatial position along the slit,
and by then integrating across the entire velocity profile of the
emission line, or by fitting one or more Gaussian profiles to the line
velocity profile.

Table ~\ref{results} summarizes the general results of our analysis that
will be described in detail below.

\subsection{Notes on individual objects}

\subsubsection{SDSS J0903+02}
The VLT and $HST$ broad band images of this QSO2 show a tadpole
morphology 
comprising a spatially asymmetric light profile, with several bright
knots along PA$\sim$110$^{\circ}$ in the central 
$\sim$3\arcsec~ (14 kpc; Fig. \ref{fig_0903}).  A low surface
brightness tail along this same 
PA extends $\sim$16\arcsec~ (75 kpc) from the position of the
quasar.  Very faint, diffuse emission is also seen to extend
$\sim$16\arcsec~ (75 kpc) to the north east.  In addition, a faint
tongue also extends $\sim$3\arcsec~ (14 kpc) to the
south east.  The narrow band image, containing [OIII] $\lambda$5007
and continuum emission, reveals a morphology that is broadly similar
to that seen in the broad band images. 

A visual inspection of our spectra reveals  spatially extended line emission along both
slit position angles (see Fig. ~\ref{fig_0903}).  This is also obvious
in Fig. ~\ref{spat} where the [OIII] spatial profile is shown in
comparison with the seeing profile. 
Along slit PA1=-65.5, the emission is dominated by a spatially
unresolved central source. In addition, we detect a compact feature of
line emitting gas in H$\alpha$, [NII], [OIII], H$\beta$ and [OII], located
1.8\arcsec~ (8 kpc) South East from the quasar.  Showing line ratios 
[OIII] $\lambda$5007 / H$\beta =$ 1.5$\pm$0.2 and [NII]$\lambda$6583
/H$\alpha =$ 0.47$\pm$0.05, this  feature falls within the `composite'
region defined by \cite{kew01} in the \cite{bal81} (BPT)
diagram. Strong underlying continuum is detected possibly associated
with this knot (although it extends further out, so this is not
certain).  
The lines are very narrow with FWHM$\la$200 km s$^{-1}$, taking into
account possible slit effects (see $\S$3.1.1).  Based on 
the compact morphology (at least along the slit), the apparent spatial
detachment from the nuclear region, the narrow lines and the composite
line ratios, we argue this is most likely a star forming object within
the quasar ionization cone. The lack of spatial information
perpendicular to the slit does not allow us 
to differentiate whether it could be a star forming companion nucleus,
a giant star forming region or even a star forming tidal feature,
spatially unresolved along the slit. 

The lines are extended on both sides of the QSO2 along slit
PA2=63.4$\degr$ (Fig. ~\ref{fig_0903}).  
Very faint (not plotted in this figure due to its weakness) [OIII]
reaches a maximum extension of  4.8\arcsec~ (22 kpc) from the 
quasar towards the West.  The [OIII]$\lambda$5007 / H$\beta$ ratio is 1.8$\pm$0.2,
indicating relatively low excitation, but insufficient by itself to
discriminate between ionization by stars or the AGN.  Because it
appears as an extended nebula connected with the QSO, we 
classify this as an extended emission line region
(EELR)\footnote{Throughout this paper, we define an EELR as a
  spatially extended nebula of line emitting gas, which shows a clear, continuous
  physical connection to the line emitting gas in the galaxy's
  nucleus.}. It shows narrow lines with FWHM$\le$180 km 
s$^{-1}$.

\begin{figure*}
\includegraphics{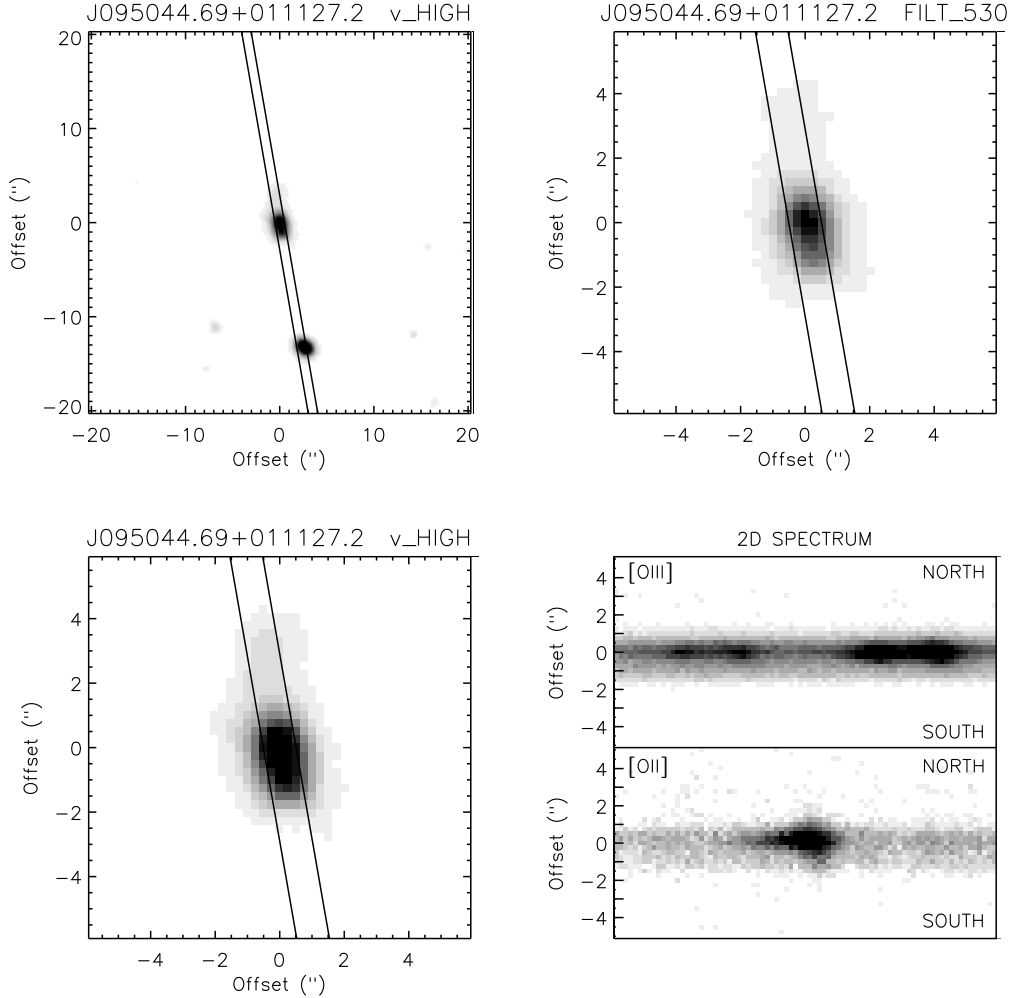}
\vspace{5.25in}
\caption{Images, the [OIII] $\lambda\lambda$4959,5007 line profile,
  and sections of the two-dimensional spectrum of SDSS 
 J095044.69+011127.2} 
\label{fig_0950}
\end{figure*}

\begin{figure*}
\includegraphics{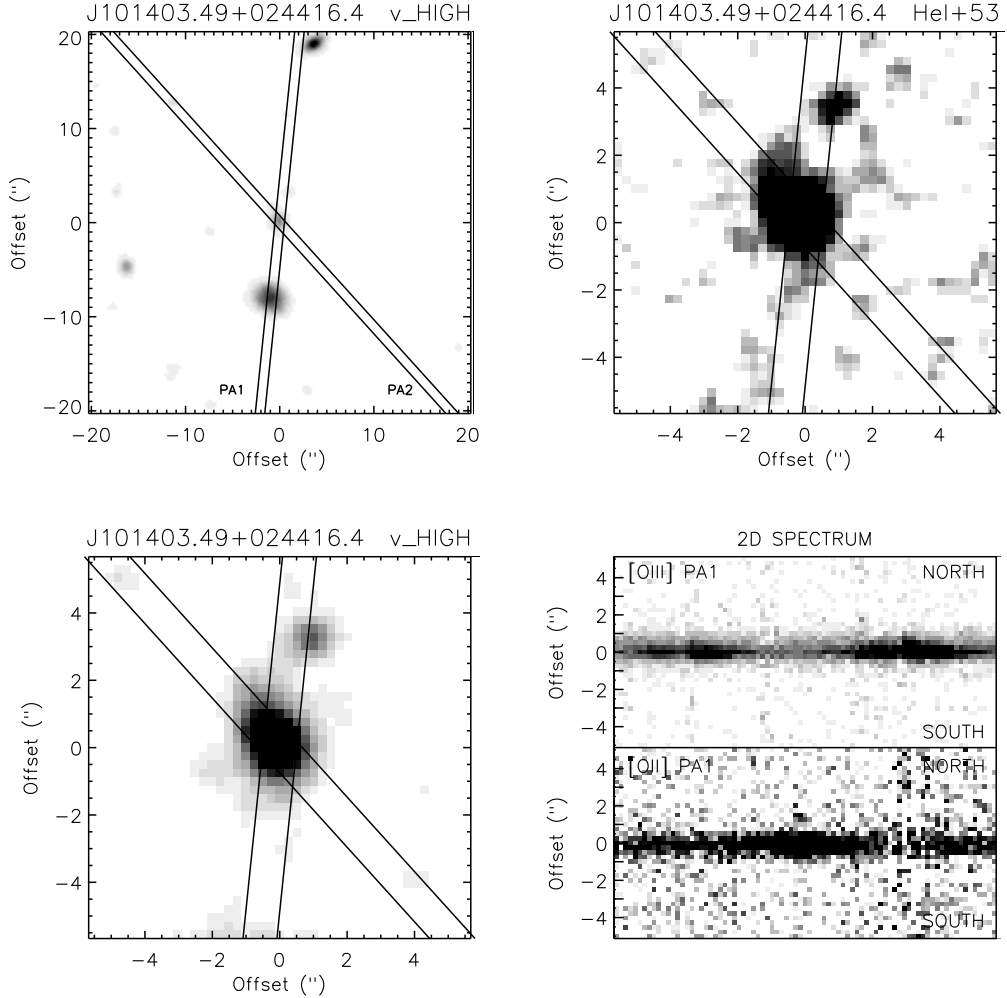}
\vspace{5.25in}
\caption{Images, the [OIII] $\lambda\lambda$4959,5007 line profile,
  and sections of the two-dimensional spectrum of SDSS 
 J101403.49+024416.4.} 
\label{fig_1014}
\end{figure*}

\begin{figure}
\includegraphics{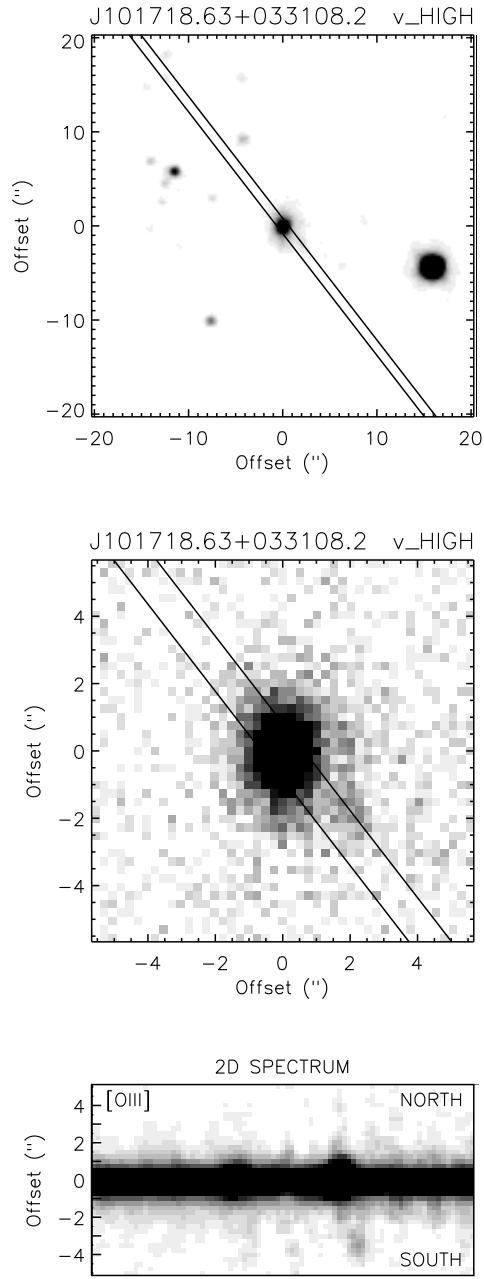}
\vspace{6.85in}
\caption{Images, the [OIII] $\lambda\lambda$4959,5007 line profile,
  and a section of the two-dimensional spectrum of SDSS 
 J101718.63+033108.2} 
\label{fig_1017}
\end{figure}

\begin{figure}
\includegraphics{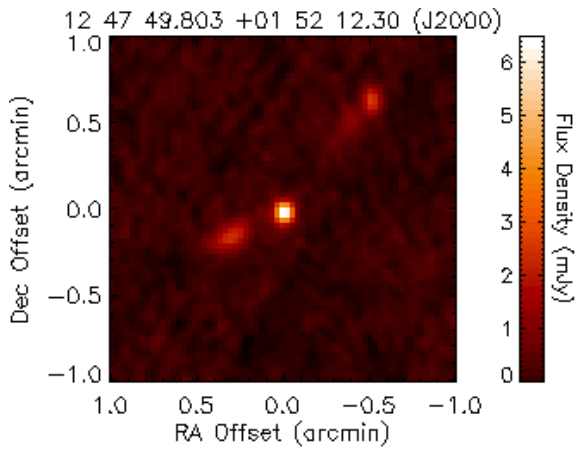}
\vspace{2.35in}
\caption{FIRST 20 cm (1.4 GHz) postage-stamp image of the radio-loud
  type 2 quasar SDSS J124749.79+015212.6. See also Lal \& Ho (2010).}
\label{first}
\end{figure}

\subsubsection{SDSS J0923+01}

\begin{figure}
\includegraphics{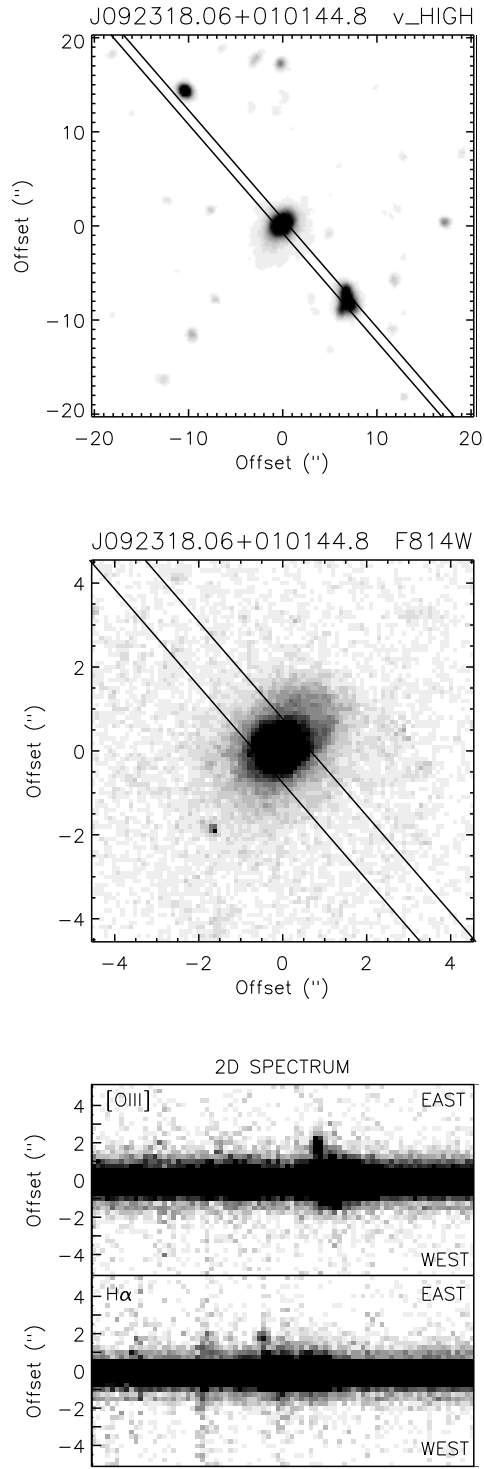}
\vspace{8in}
\caption{Images, the [OIII] $\lambda\lambda$4959,5007 line profile,
  and sections of the two-dimensional spectrum of SDSS
  J092318.06+010144.8} 
\label{fig_0923}
\end{figure}

The broad band morphology of the host galaxy of this QSO2 was described in detail by
\cite{bes12}; based on deep Gemini-GMOS  broad band optical images, they
found a strong morphological disturbance. Our FORS2 optical image shows
a galaxy morphology that is elongated along a 
PA$\sim$-45$^{\circ}$, with low surface brightness features that resemble
tails and broad fans (Fig. \ref{fig_0923}). In addition, the central few arcsec shows a lopsided
flux distribution. 

Modeling of the stellar populations in the host galaxy of this QSO2
reveals a large contribution from a young stellar population with an
age of between 50 and 100 Myr (Bessiere et al. 2015, in prep.).

The spatial profile of the emission lines along the slit
PA=40.9$^{\circ}$ is dominated by a spatially unresolved central
component (Fig. \ref{spat}) with FWHM$\sim$0.61\arcsec, consistent within the errors with the seeing   
values derived from  broad band image (0.66$\pm$0.06\arcsec; Table  \,\ref{tab:obs}).
 In addition, both [OIII]$\lambda$5007 and H$\alpha$ show a faint and compact  (along the slit)  feature of
narrow emission lines (FWHM$<$200 km s$^{-1}$, taking into account possible slit effects). It is located at 2\arcsec~ (10 kpc) north east of the AGN, blueshifted by 550 km s$^{-1}$ 
(ignoring slit effects)  relative to the nuclear emission.  Continuum is not detected at the location of the knot.  The object is clearly seen in the two-dimensional spectrum  (Fig. \ref{fig_0903}) and also as a faint excess above the seeing profile in Fig. \ref{spat}.

This knot shows [OIII] $\lambda$5007 / H$\beta$
$\ga$6 and [NII] / H$\alpha$ $\la$0.2 ratios that are suggestive of 
stellar photoionization (e.g., Kewley et al. \citeyear{kew01}). The
compact appearance along the slit, the narrow lines and the line ratios
suggest that this is a star forming object. 

Also within the slit is the unrelated irregular galaxy SDSS 
J092317.57+010136.6 (see Fig. ~\ref{fig_0923}) at $\sim$11\arcsec to the SW.
The  detected [OIII]
$\lambda\lambda$4959,5007 and H$\alpha$ lines place it at z$=$0.191.

\subsubsection{SDSS J0950+01}

This HLSy2 shows a strongly asymmetric light profile in its broad
band image (Fig. \ref{fig_0950}).  A low surface brightness tail extends
$\sim$4\arcsec~ (22 kpc) North from the quasar, along a similar position angle
to that of the high surface brightness regions of the galaxy, and a
faint tail (or possibly a poorly resolved companion nucleus) also extends $\sim$5\arcsec~ (27 kpc) 
southward.  The narrow band
image, which samples [OII] $\lambda$3727 and continuum emission, shows 
similar morphological structure. 

[OII] is the only line confirmed to be spatially extended, while the [OIII] profile is consistent with a seeing disk (Fig. ~\ref{spat}) of FWHM$\sim$1.0\arcsec. 
Seeing broadening at decreasing $\lambda$ (expected to be $\le$16\%,
see \S3.1.1) cannot explain this difference in the spatial profiles of
[OII] and [OIII].
The [OII] spatial profile suggests that the line is emitted by an EELR associated with the HLSy2. The total
observed extent is  4.5\arcsec~ (24 kpc) and maximum extent from the continuum centroid
of 2.3\arcsec~(12 kpc; Table~\ref{results}). The line is very narrow
in the extended gas, with a total doublet FWHM$\la$235 km s$^{-1}$ and 215$\pm$45 km s$^{-1}$  
respectively at both sides of the continuum centroid, taking possible
slit effects into account.  

\subsubsection{SDSS J1014+02}

In the broad band image, the host galaxy of this HLSy2 shows a
cometary, asymmetrical light profile in its higher surface brightness
region, with an arc (or tail) of low surface brightness emission
extending North East and then North West, terminating in a knot of
emission $\sim$3.25\arcsec~ (21 kpc) North of the galaxy
(Fig. ~\ref{fig_1014}). 

Our spectroscopic data shows  that the [OIII] spatial profiles along
both position angles  are dominated by the  central spatially
unresolved source. Along PA1=-5.9, this  appears somewhat narrower
than the seeing derived from the narrow band image
(FWHM$\sim$0.64$\pm$0.03 \arcsec) (Fig. ~\ref{spat}). The slightly
broader central [OII] spatial profile along PA1 is consistent with
seeing broadening at shorter $\lambda$. Along PA2, [OII] appears
slightly extended. 

In addition, very low surface brightness (too faint to be plotted in
Fig. ~\ref{spat}), diffuse [OII] is detected  along PA2  towards the
South-West up to $\sim$2.8\arcsec or 18 kpc from the continuum
centroid. The lines are very
narrow in this EELR. The doublet has FWHM$\la$140 km s$^{-1}$ in the
EELR -- in the most extreme case of maximum possible slit effects, the
lines could have at most FWHM=211$\pm$37 km s$^{-1}$. 

The galaxy 43\arcsec~ North East from the AGN along PA2
detected in extended [OII] emission with z$=$0.571, confirming its
association with the quasar.  The galaxies 8\arcsec~ south and 19\arcsec~ north along
PA1  unrelated star forming galaxies at
substantially different redshifts.  

\subsubsection{SDSS J1017+03}

The broad band image of this HLSy2 shows an extended  low surface
brightness amorphous halo and a compact knot located 
3\arcsec~ (17 kpc)  SW of the nucleus (Fig.~\ref{fig_1017}).

The [OIII] $\lambda$5007 spatial profile is dominated by a central
unresolved source (Fig.~\ref{spat}). A faint excess above the seeing
profile is detected at $\sim$1.2\arcsec from the spatial centroid
towards the NE. In addition,  
faint spatially extended emission is detected  toward the SW
(too faint to be discerned in Fig.~\ref{spat}). It reaches a distance
of 3.5\arcsec~ (20 kpc) from the AGN (Fig.~\ref{fig_1017}) where it
seems to connect with the position of the knot seen in the broad band
image. This knot also shows line emission.  The [OIII] line is very
narrow with FWHM$\la$140 km s$^{-1}$ (309$\pm$30 accounting for the
most extreme possible slit effects)  along its full extension and it is
redshifted by up to $\sim$530 km s$^{-1}$  
relative to the [OIII] emission at the position of peak flux. For
the [OIII] $\lambda$5007 / H$\beta$  ratio we obtain a 3$\sigma$
lower limit of $\ge$4.

\subsubsection{SDSS J1247+01}
\begin{figure*}
\includegraphics{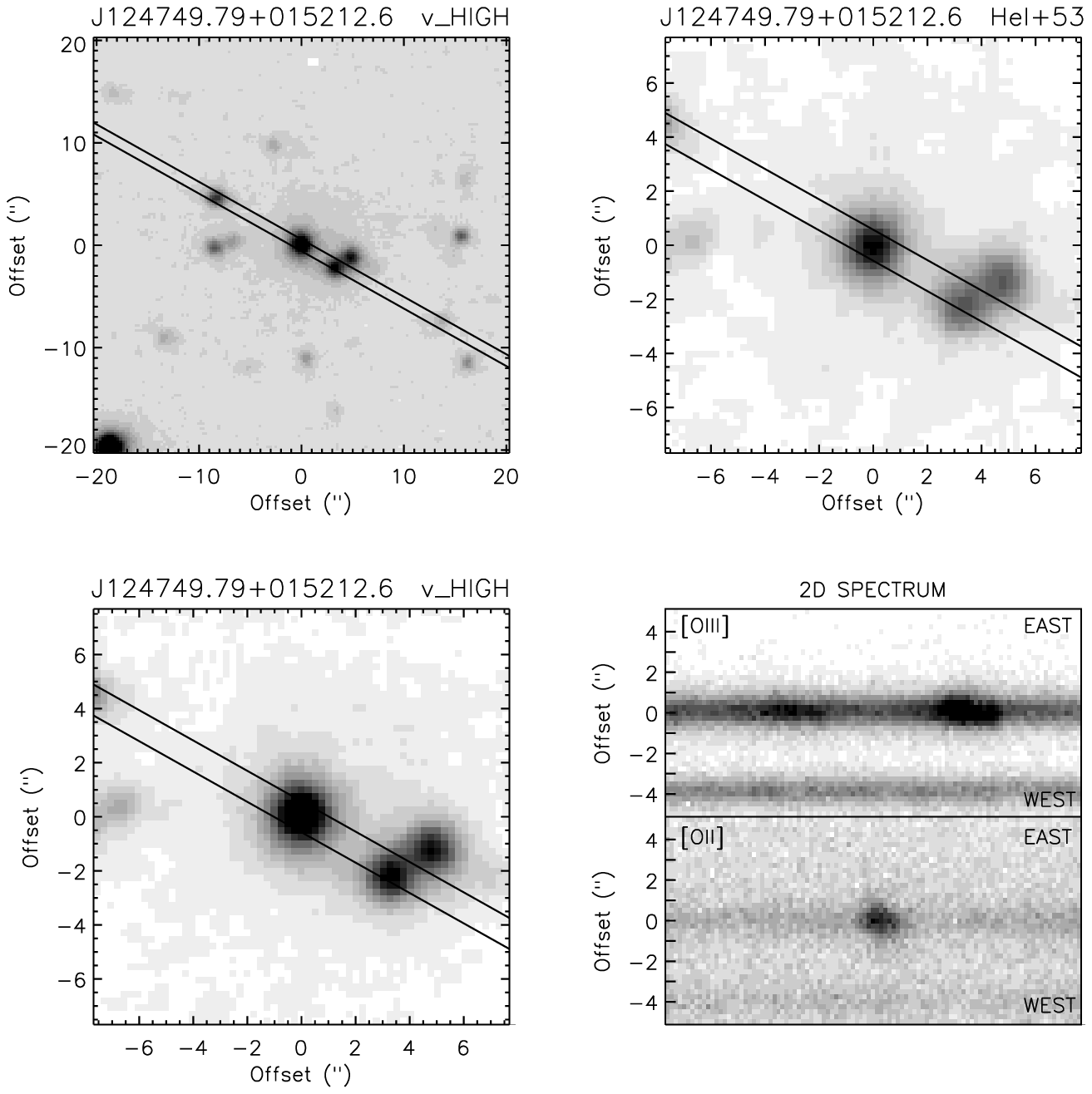}
\includegraphics{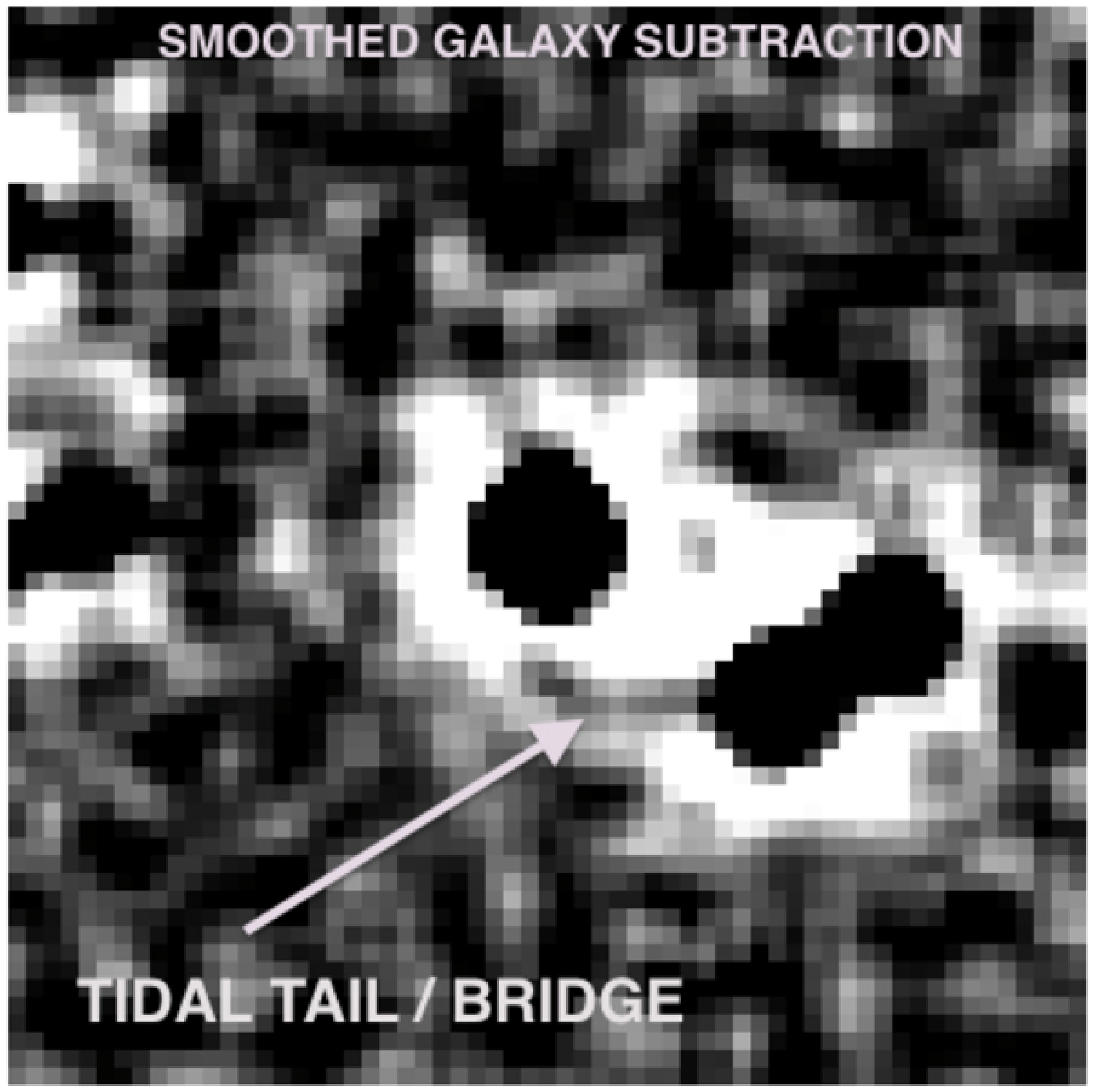}
\vspace{8.4in}
\caption{Images and the [OIII] $\lambda\lambda$4959,5007 line profile,
  of SDSS J124749.79+015212.6. The lower panel shows the central
  15.25\arcsec$\times$15.25\arcsec of broad band
  image of the field around the HLSy2, after applying the
  'smoothed galaxy subtraction' technique of Ramos Almeida et
  al. (2011). This analysis has revealed a tidal tail (or bridge) connecting the
  HLSy2 to its nearest companion galaxy, visible as a dark arc to the
  south and south-west of the HLSy2.}  
\label{fig_1247}
\end{figure*}

The broad band image shows several other galaxies within 10\arcsec~
(56 kpc) of
the HLSy2, including a pair of galaxies located 4\arcsec~
(22 kpc) 
south west of the HLSy2 (Fig.~\ref{fig_1247}).  One member of 
this galaxy pair fell within the slit, and shows weak [OII]
emission and a continuum break consistent with the 4000 \AA~ break at
z=0.42, similar to that of the AGN host. 

We detect a faint tidal tail extending $\sim$2\arcsec~ west from the 
western-most of the companion galaxy pair. In addition, after processing the
broad band image using the 'smoothed galaxy subtraction' technique of
Ramos Almeida et al. (2011), we detect a tidal tail (or bridge)
connecting the HLSy2 to its nearest companion galaxy (see lower panel
in Fig.~\ref{fig_1247}). We conclude that this is a triple
merger/interacting system.

Our spectroscopic data shows that the [OII] and [OIII] spatial profiles along PA=60.7 are both spatially extended. The [OII] and
[OIII] profiles are shown in Fig. ~\ref{spat} together with a seeing
profile of FWHM$\sim$1.2\arcsec.   Both [OII] and [OIII] show a clear excess above 
the seeing disk at both sides of the spatial centroid. This is also clear in the 2-dim spectrum (Fig.~\ref{fig_1247}). A visual inspection shows that  extended  faint [OII]  is detected 
 along the slit  up to $\sim$3\arcsec NE from the continuum centroid ($\sim$17 kpc), with well differentiated kinematics.
 The lines appear spectrally unresolved with FWHM$\la$120 km s$^{-1}$
 (slit effects do not affect this object).
For [OIII], the excess is clear towards the east. 

  This galaxy was detected by the FIRST survey (20 cm / 1.4 GHz) with a flux density of
7.1$\pm$0.1 mJy. In addition to a bright radio core, the FIRST image
also shows two diametrically opposed radio lobes with a separation of
$\sim$70\arcsec (400 kpc; Fig. ~\ref{first}). The substantially higher NVSS flux density
of 34.1$\pm$1.6 mJy (20 cm / 1.4 GHz) further attests to the presence of large-scale,
low surface brightness radio emission (see also Lal \& Ho 2010).

The galaxy at $\sim$9\arcsec to the North East from the AGN has 
photometric $z=$0.456$\pm$0.026, according to the SDSS
information. This suggests that it could also be associated with the
quasar, but spectroscopic z confirmation is needed.

\subsubsection{SDSS J1336-00}
\begin{figure}
\includegraphics{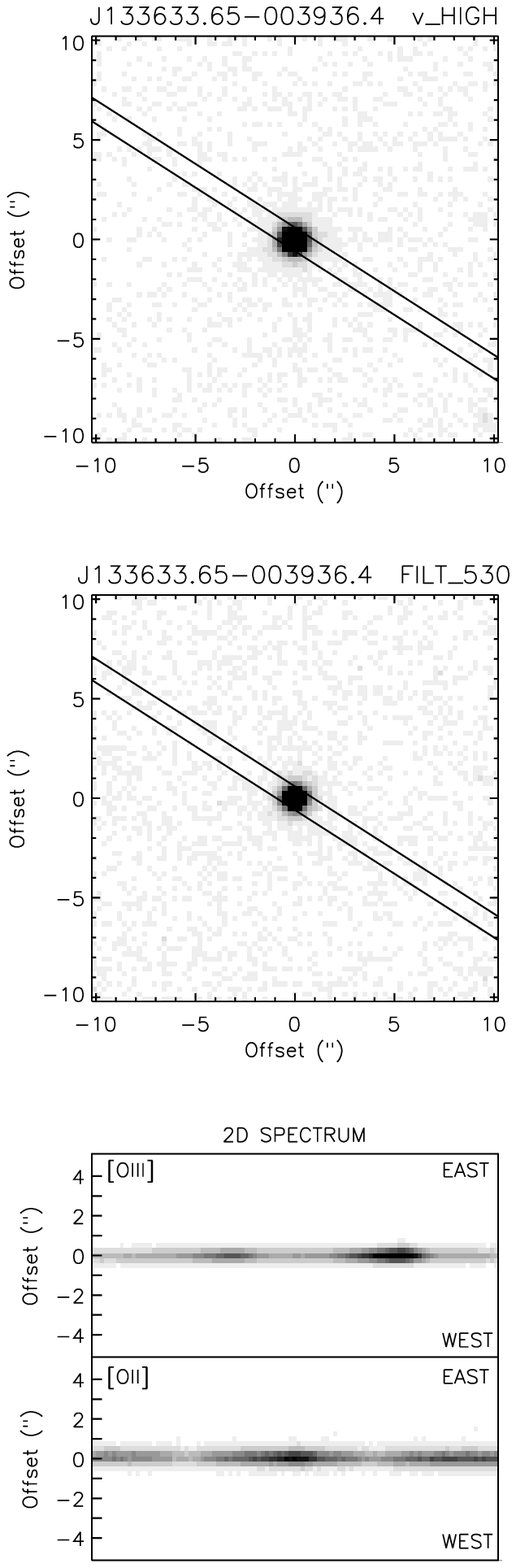}
\vspace{8in}
\caption{Images and the [OIII] $\lambda\lambda$4959,5007 line profile
  of SDSS J133633.65-003936.4.}
\label{fig_1336}
\end{figure}

\begin{figure}
\includegraphics{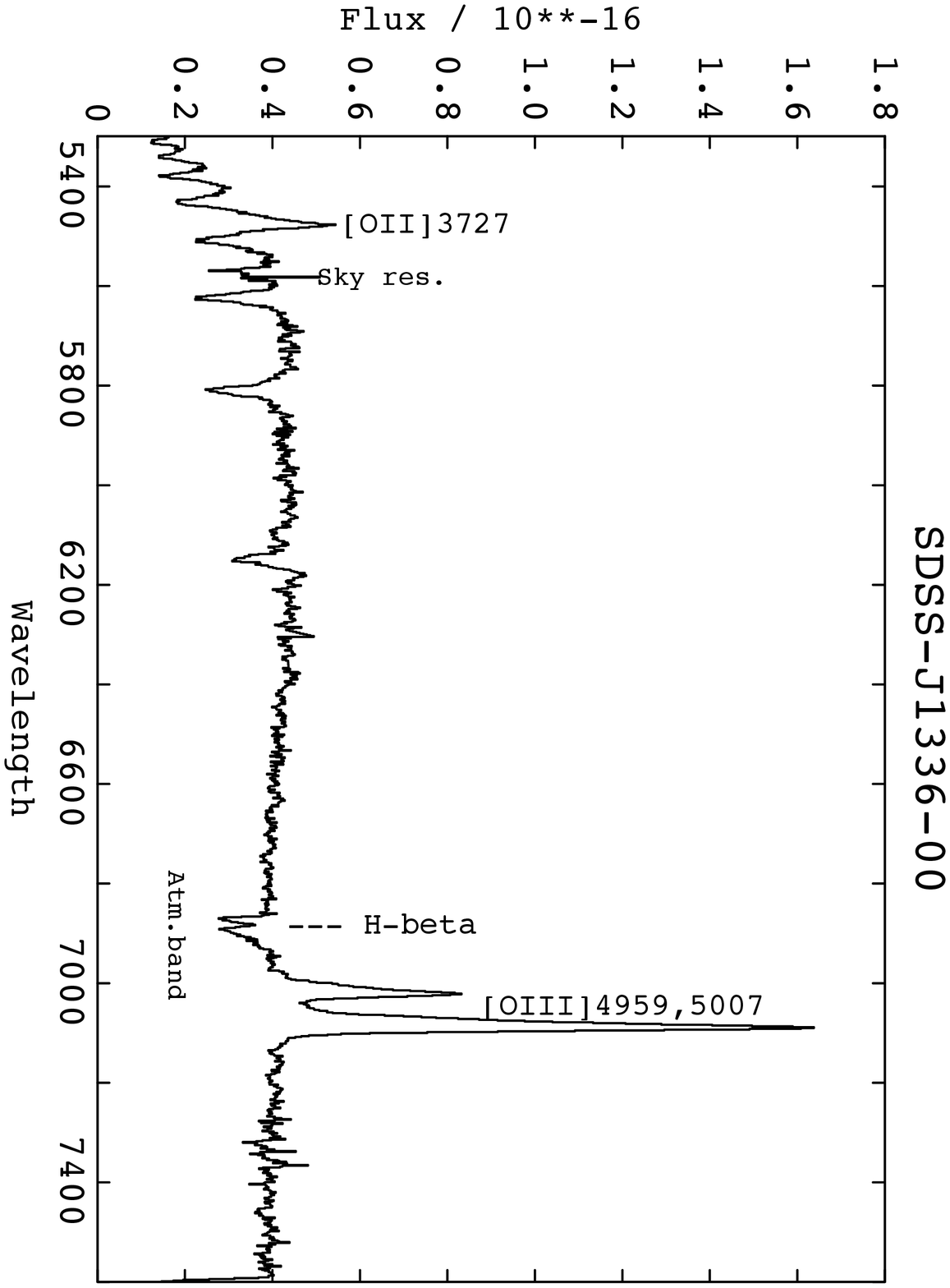}
\vspace{2.5in}
\caption{Nuclear spectrum of SDSS J1336-00. Strong Balmer absorption lines and the Balmer break suggest the presence of a post-starburst
stellar population of $\ga$100 Myr of age.  H$\beta$ is affected by an atmospheric band. The flux is in units of $\times$10$^{-16}$ erg s$^{-1}$ cm$^{-2}$ arcsec$^{-2}$.}
\label{nuc1336}
\end{figure}

This apparently isolated galaxy shows an  elliptical compact morphology in our broad
and intermediate band images (Fig. ~\ref{fig_1336}).  No spatial
structure is apparent, and 
there are no other bright galaxies visible within a radius of at least
80 kpc.  In spite of this unremarkable appearance, the optical spectrum shows strong Balmer absorption lines and  a strong Balmer break
 (Fig. ~\ref{nuc1336}), suggesting a burst of star formation occurred $\ga$ 100 Myr ago. 

\begin{figure*}
\includegraphics{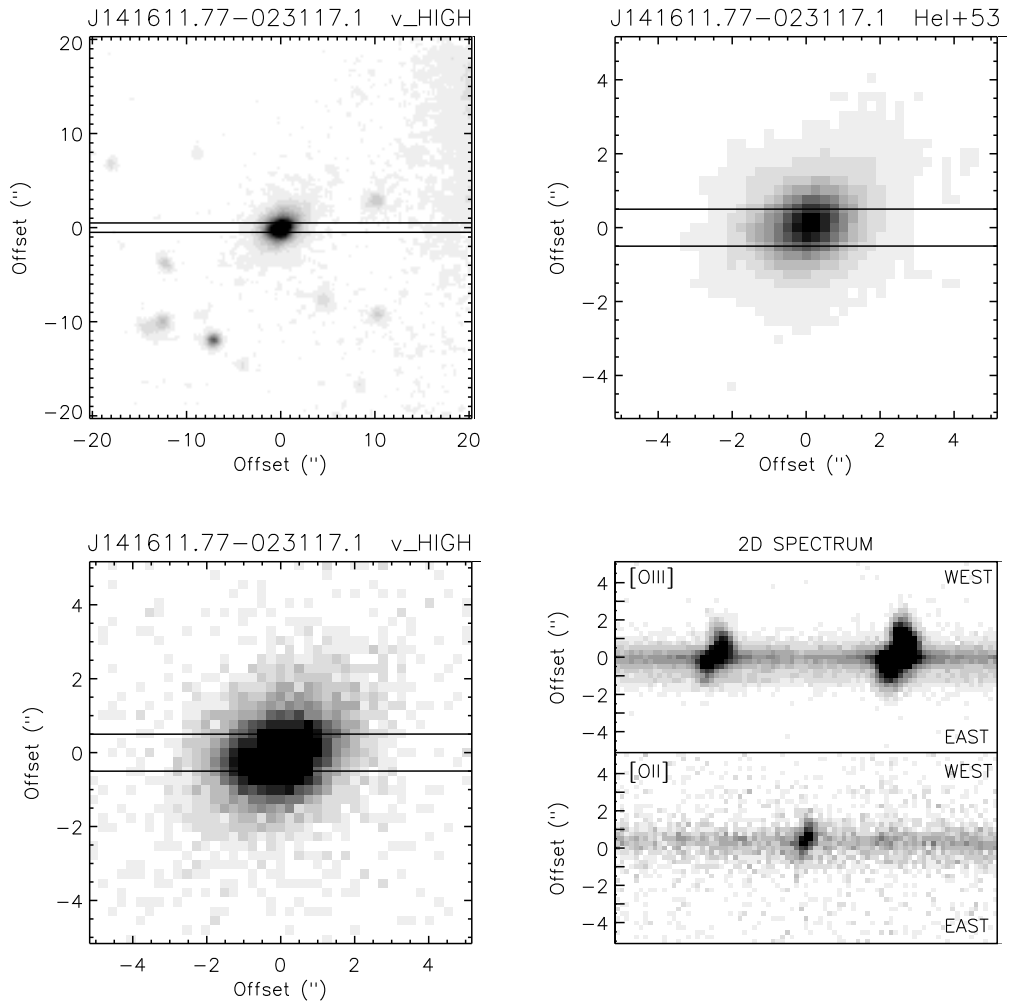}
\vspace{5.25in}
\caption{Images, the [OIII] $\lambda\lambda$4959,5007 line profile and
  sections of the two-dimensional spectrum
  of SDSS J141611.77-023117.1.}
\label{fig_1416}
\end{figure*}

\begin{figure}
\includegraphics{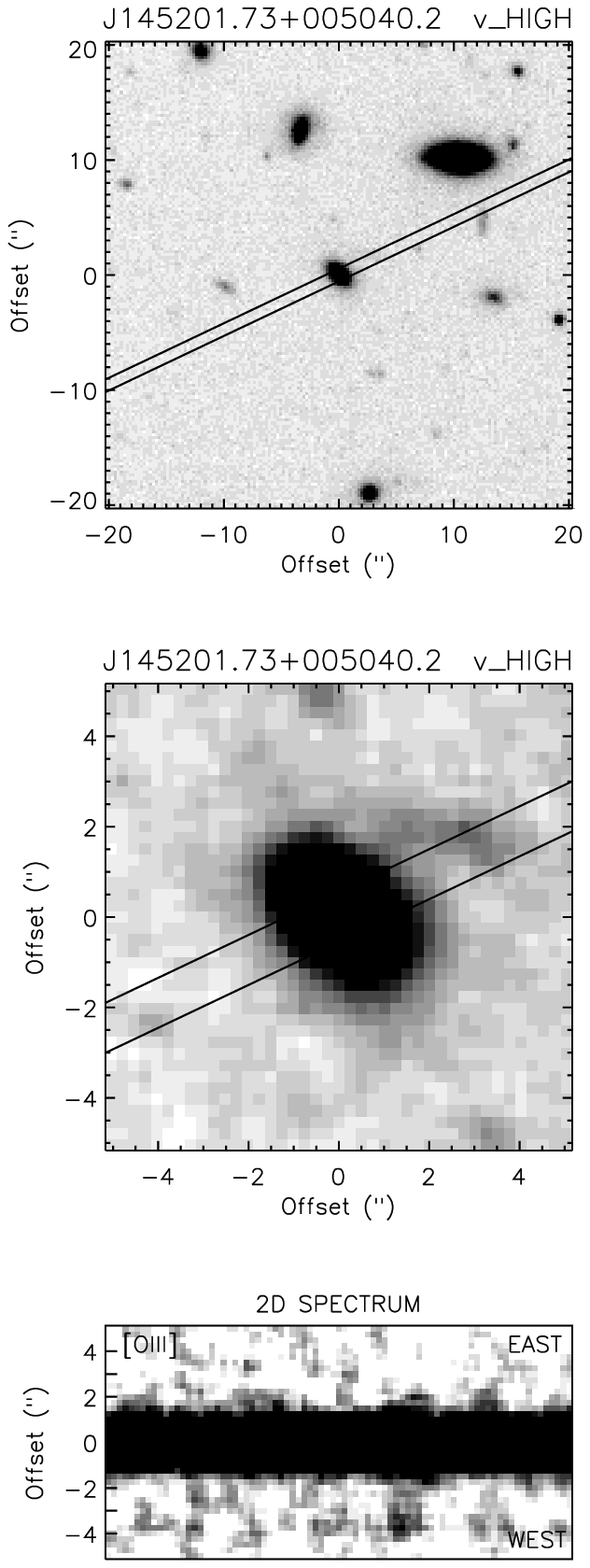}
\vspace{6.3in}
\caption{Images, the [OIII] $\lambda\lambda$4959,5007 line profile and
  a section of the two-dimensional spectrum
  of SDSS J145201.73+005040.2.}
\label{fig_1452}
\end{figure}

\begin{figure*}
\includegraphics{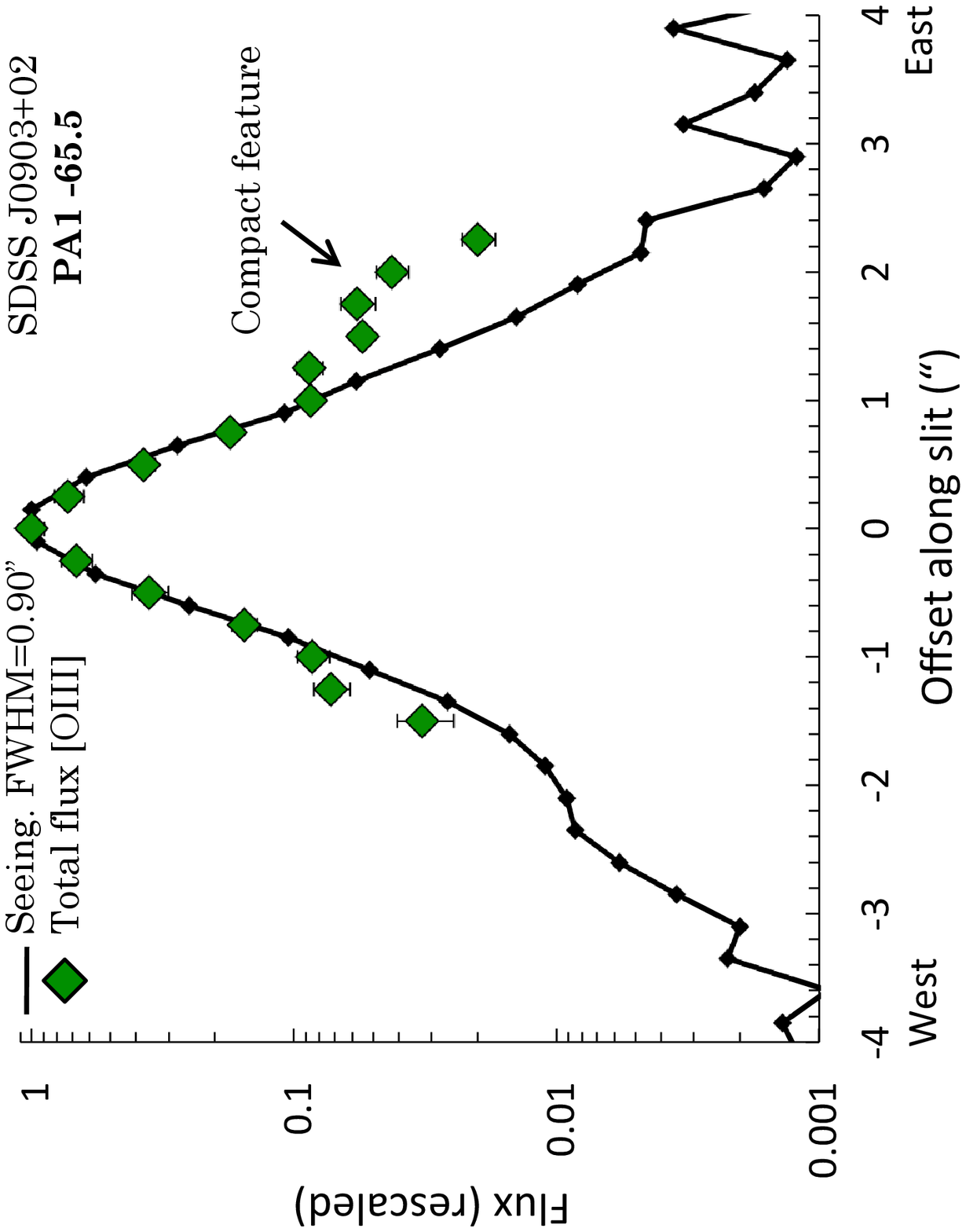}
\includegraphics{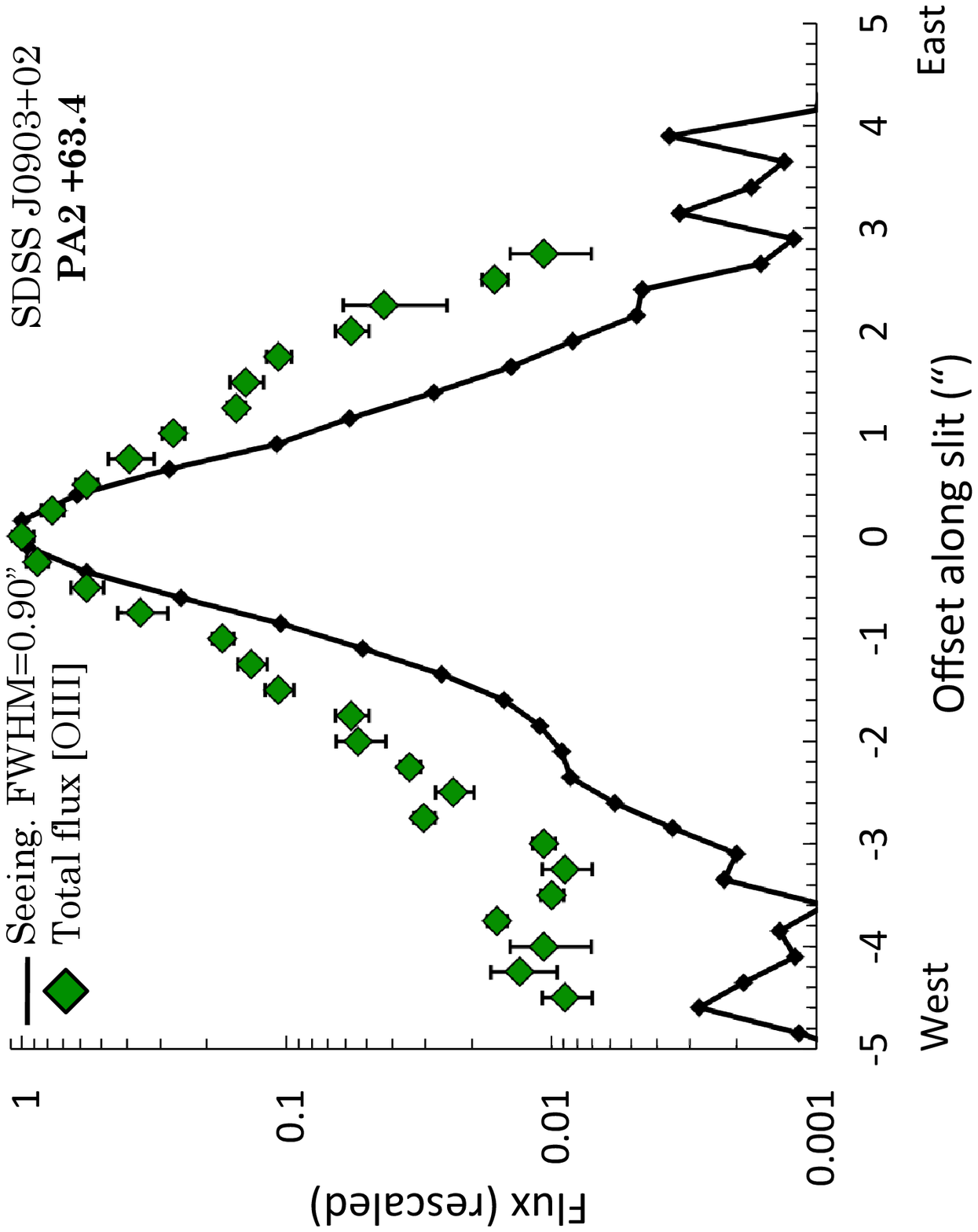}
\includegraphics{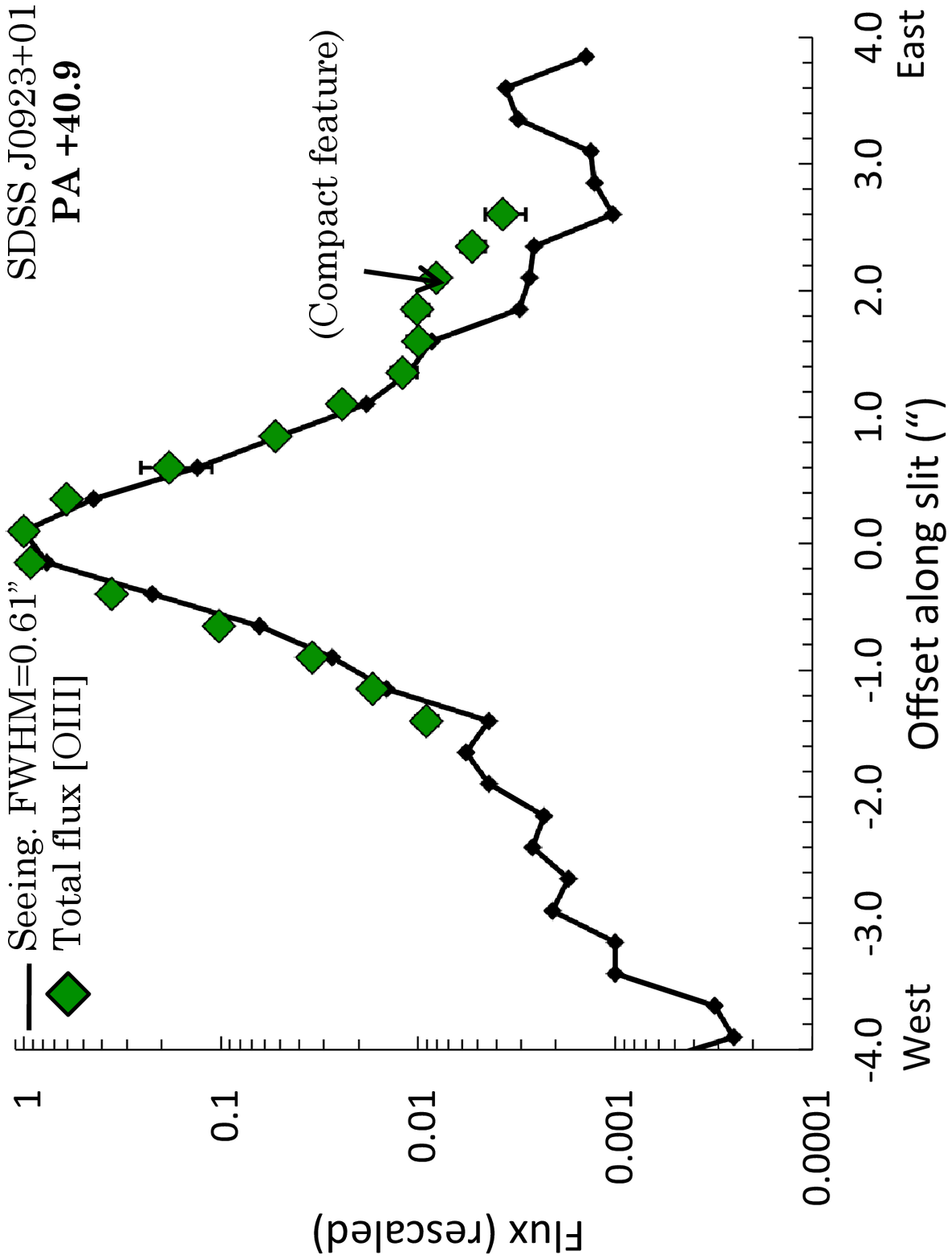}

\includegraphics{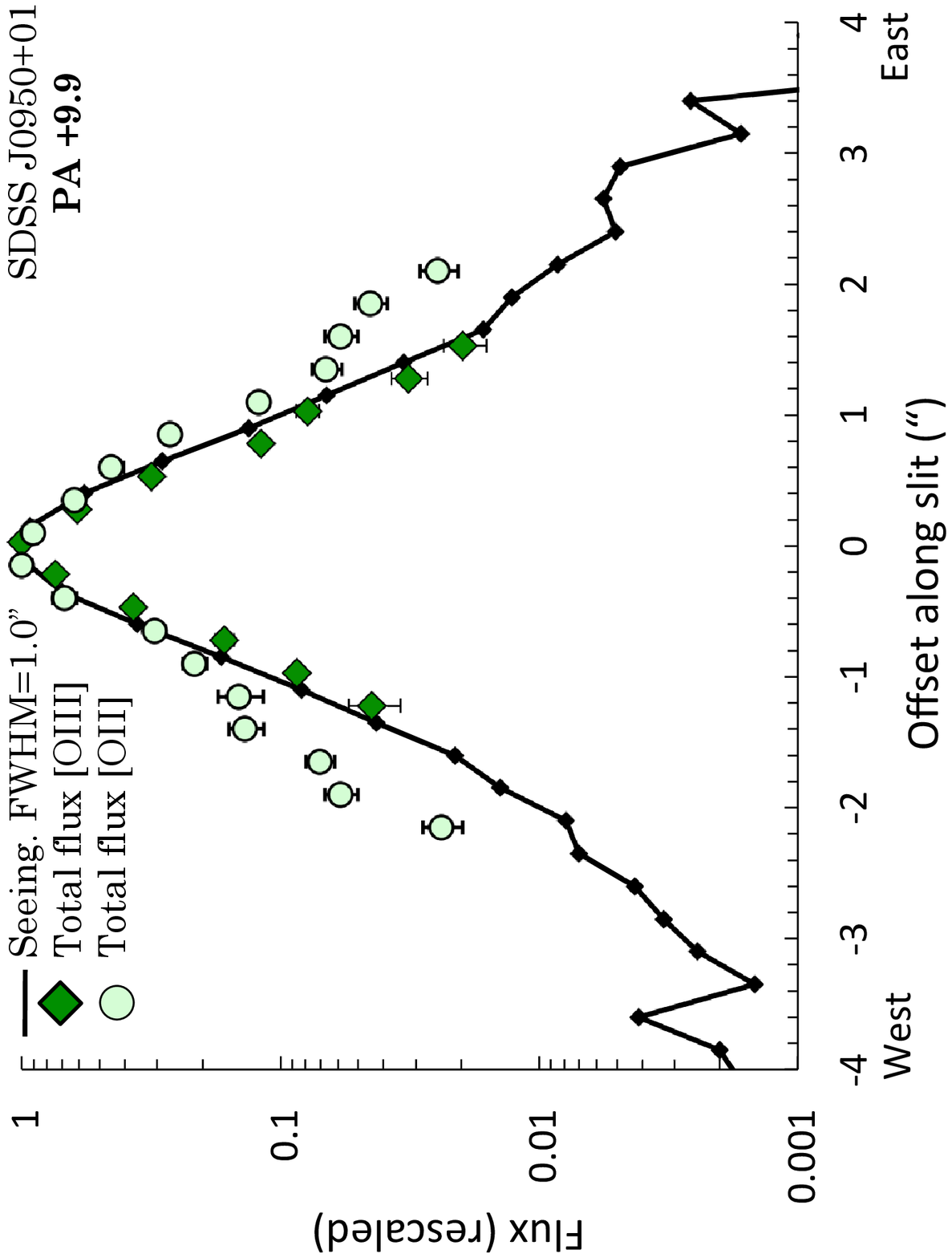}
\includegraphics{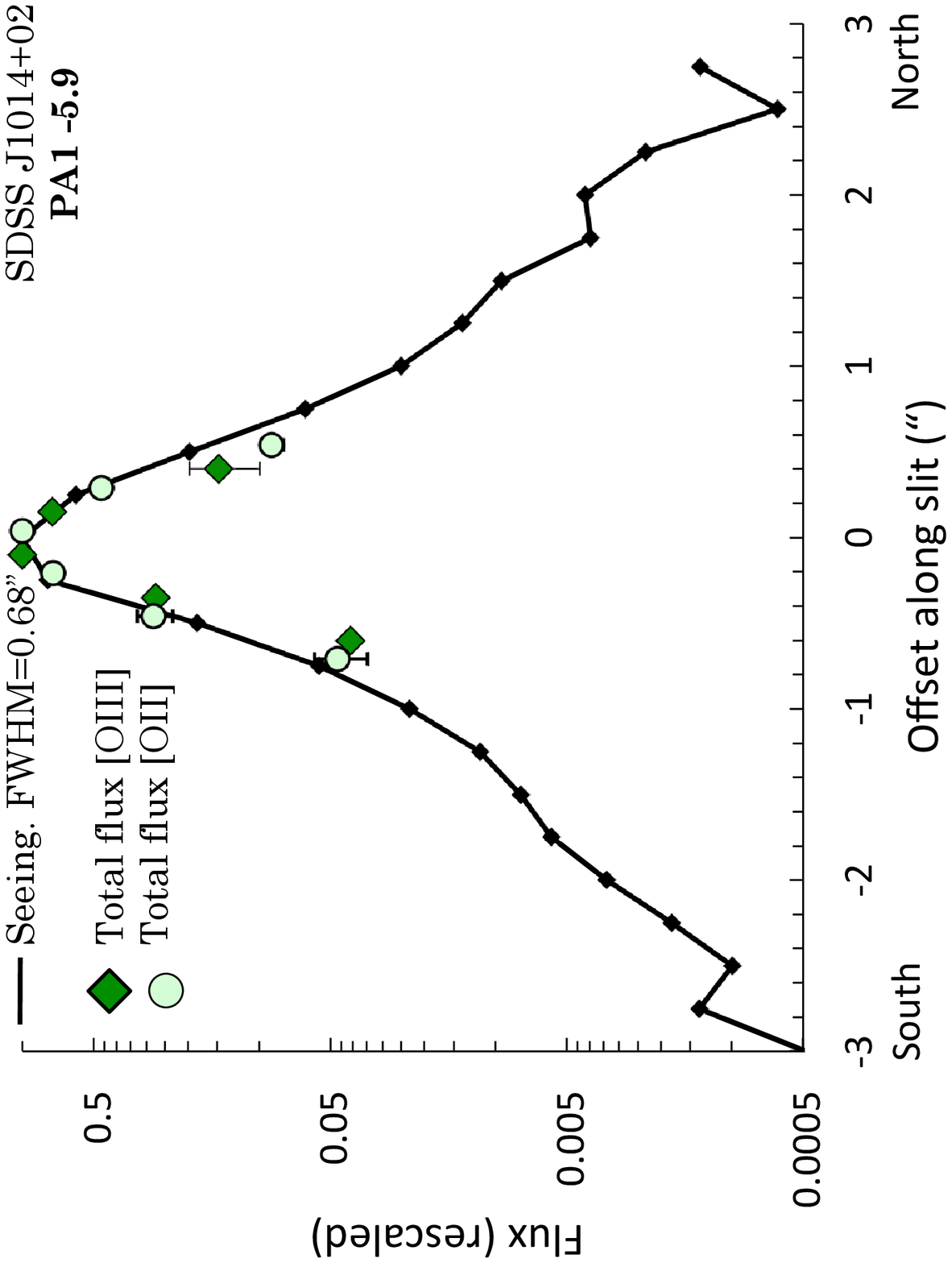}
\includegraphics{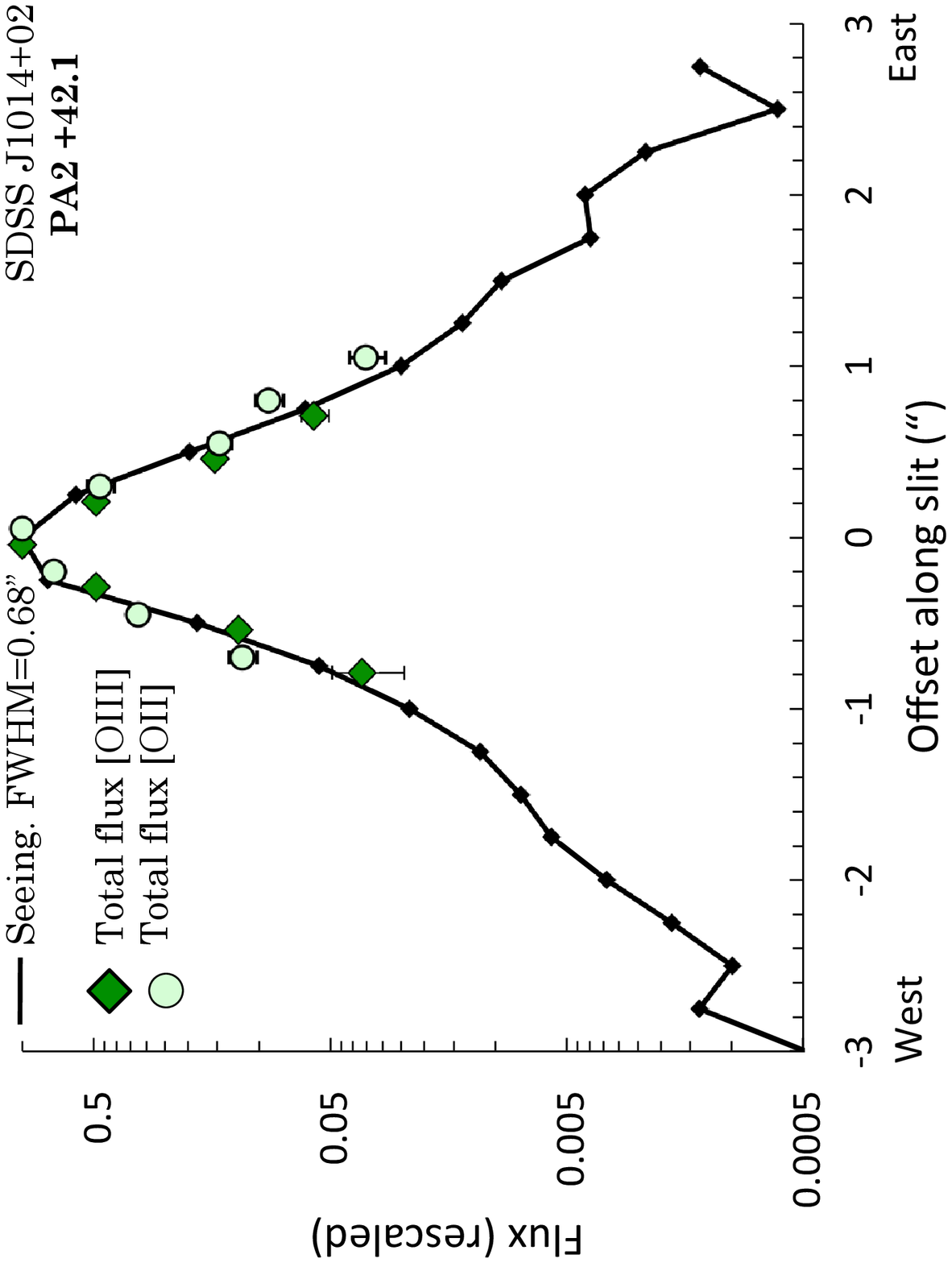}

\includegraphics{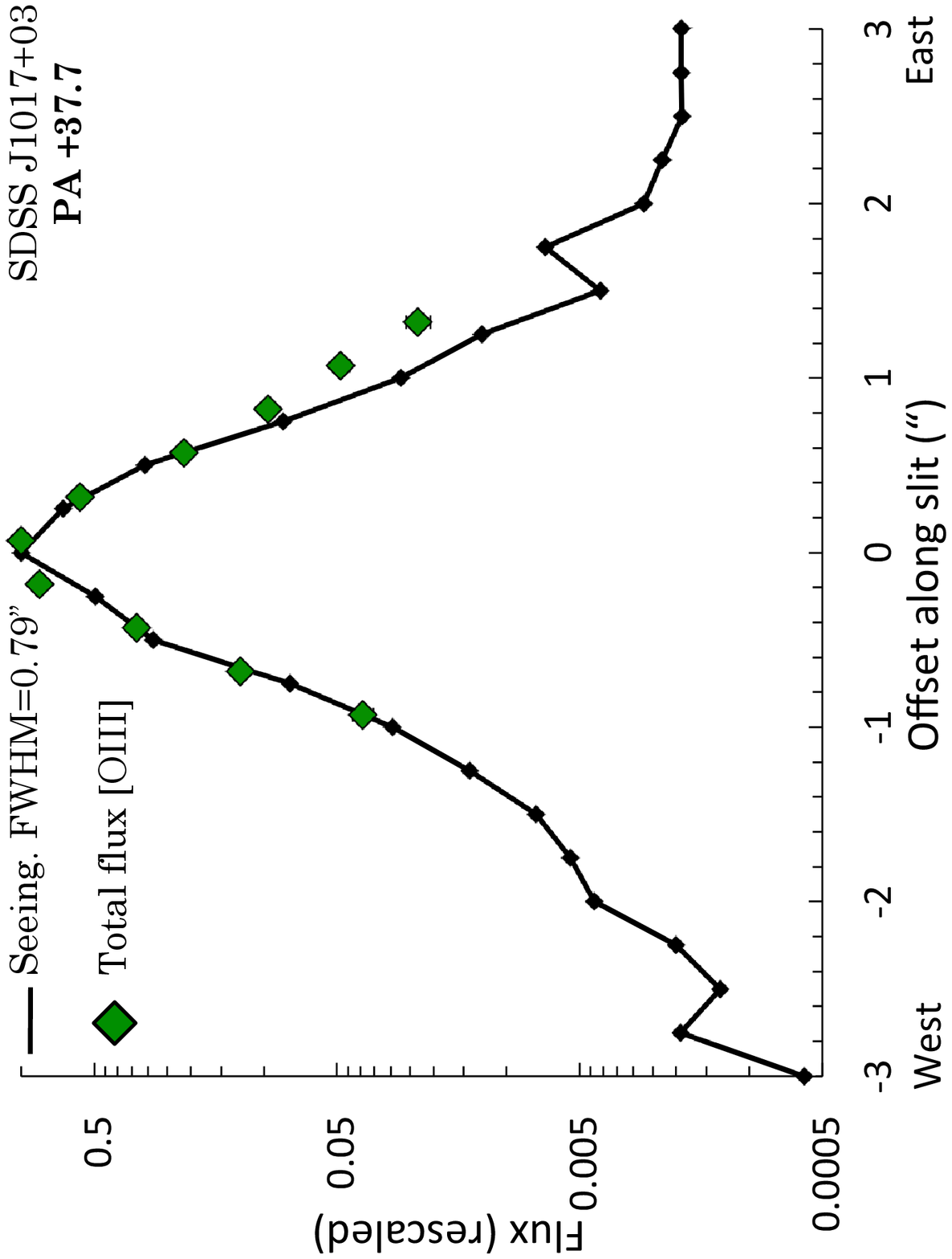}
\includegraphics{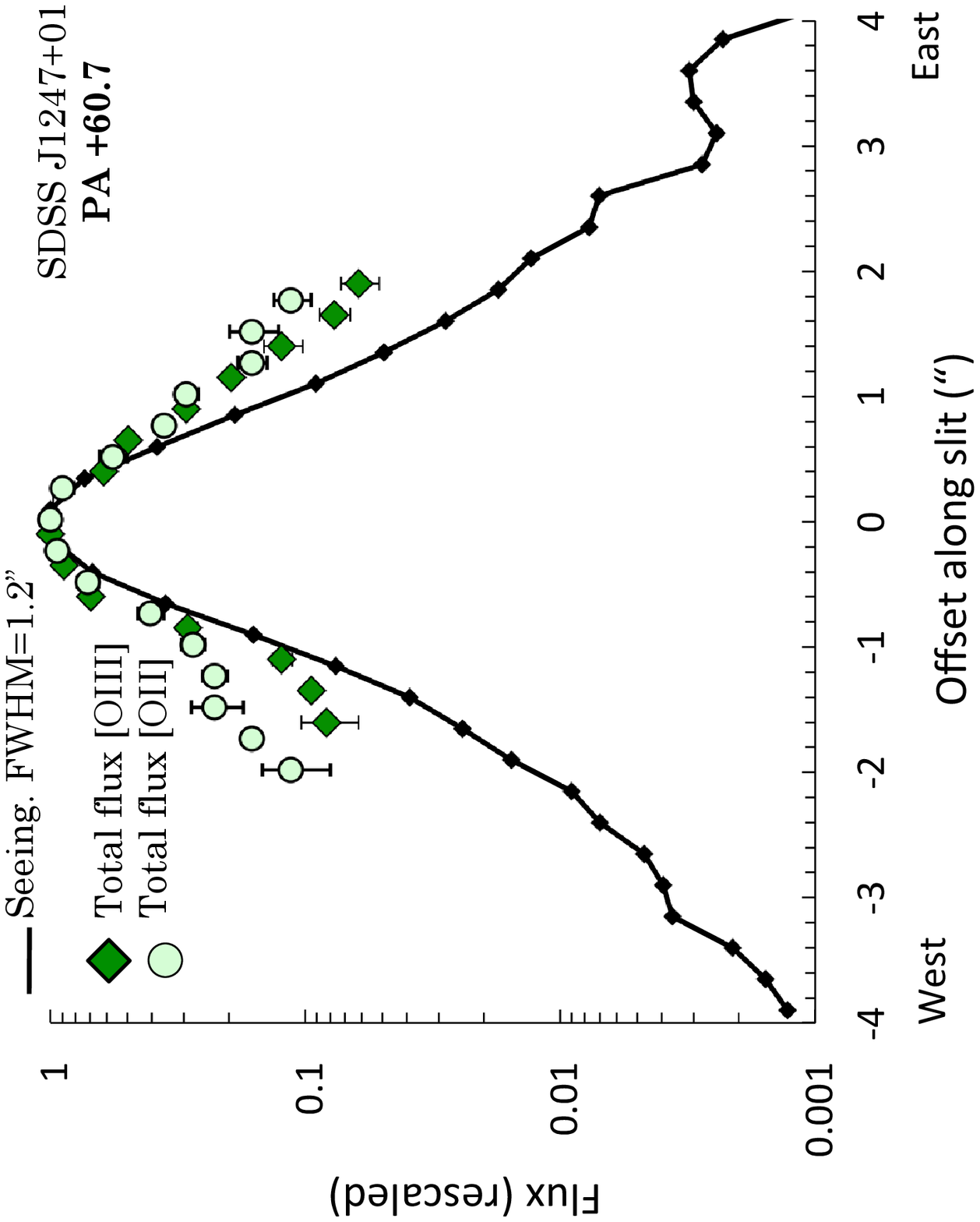}
\includegraphics{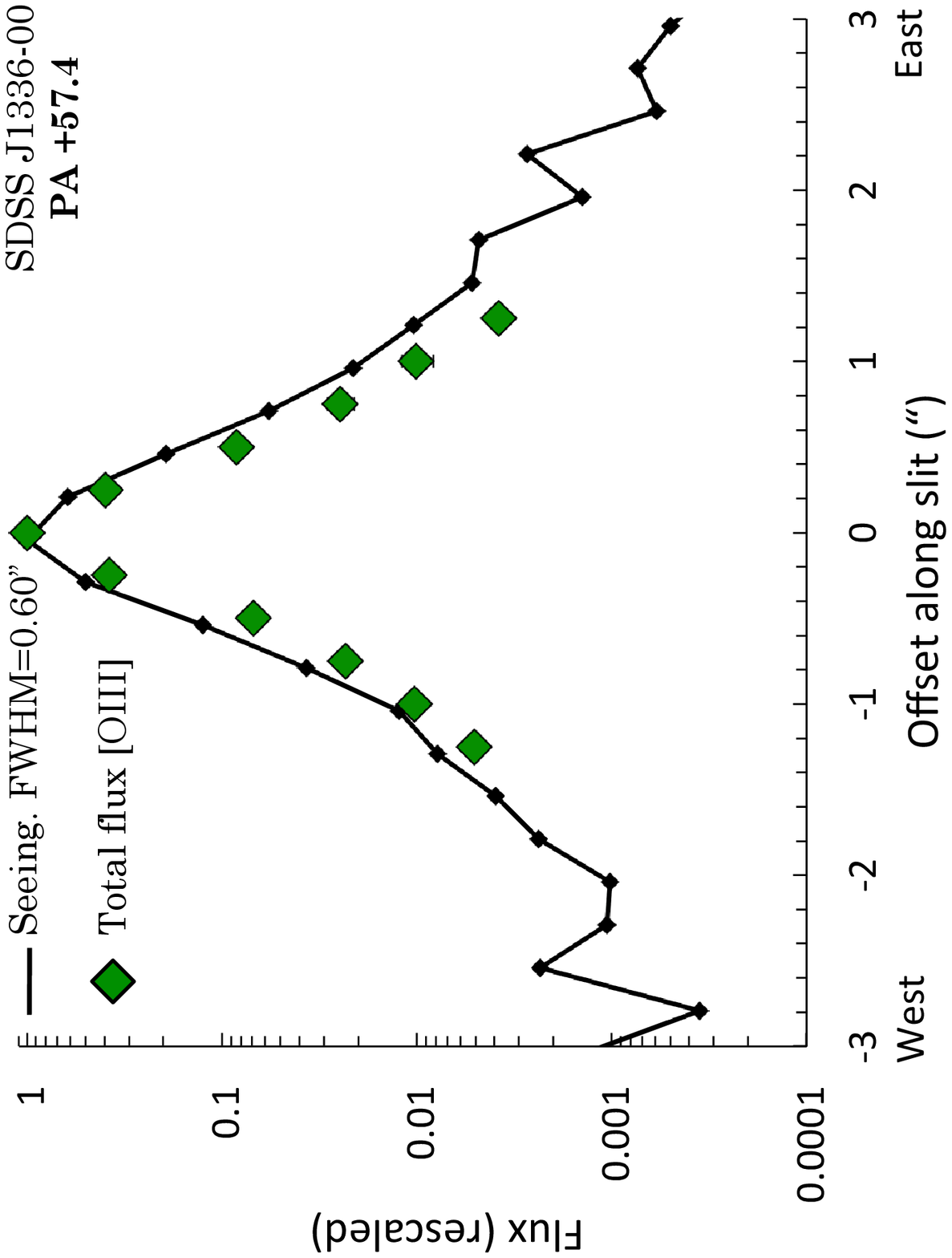}

\includegraphics{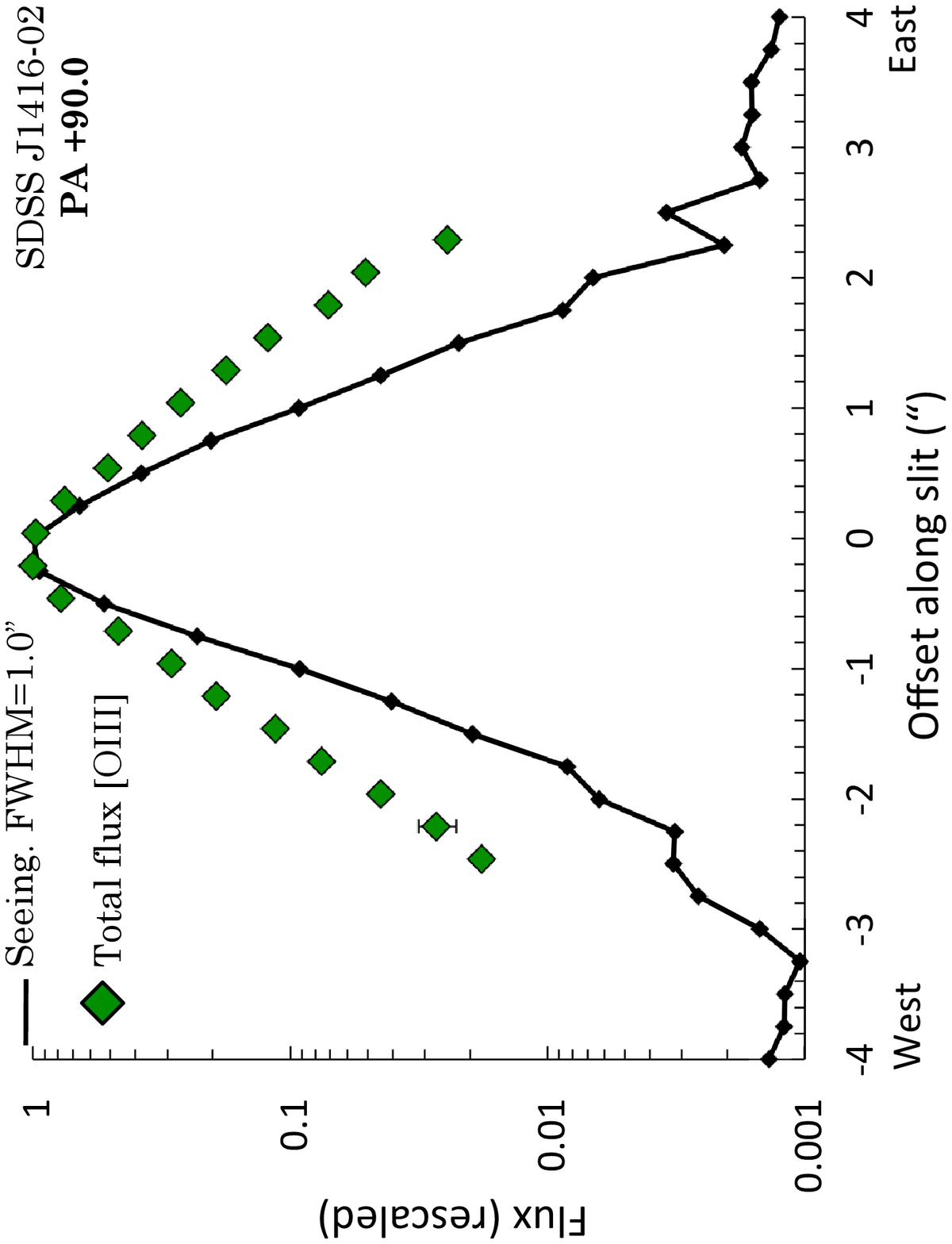}
\includegraphics{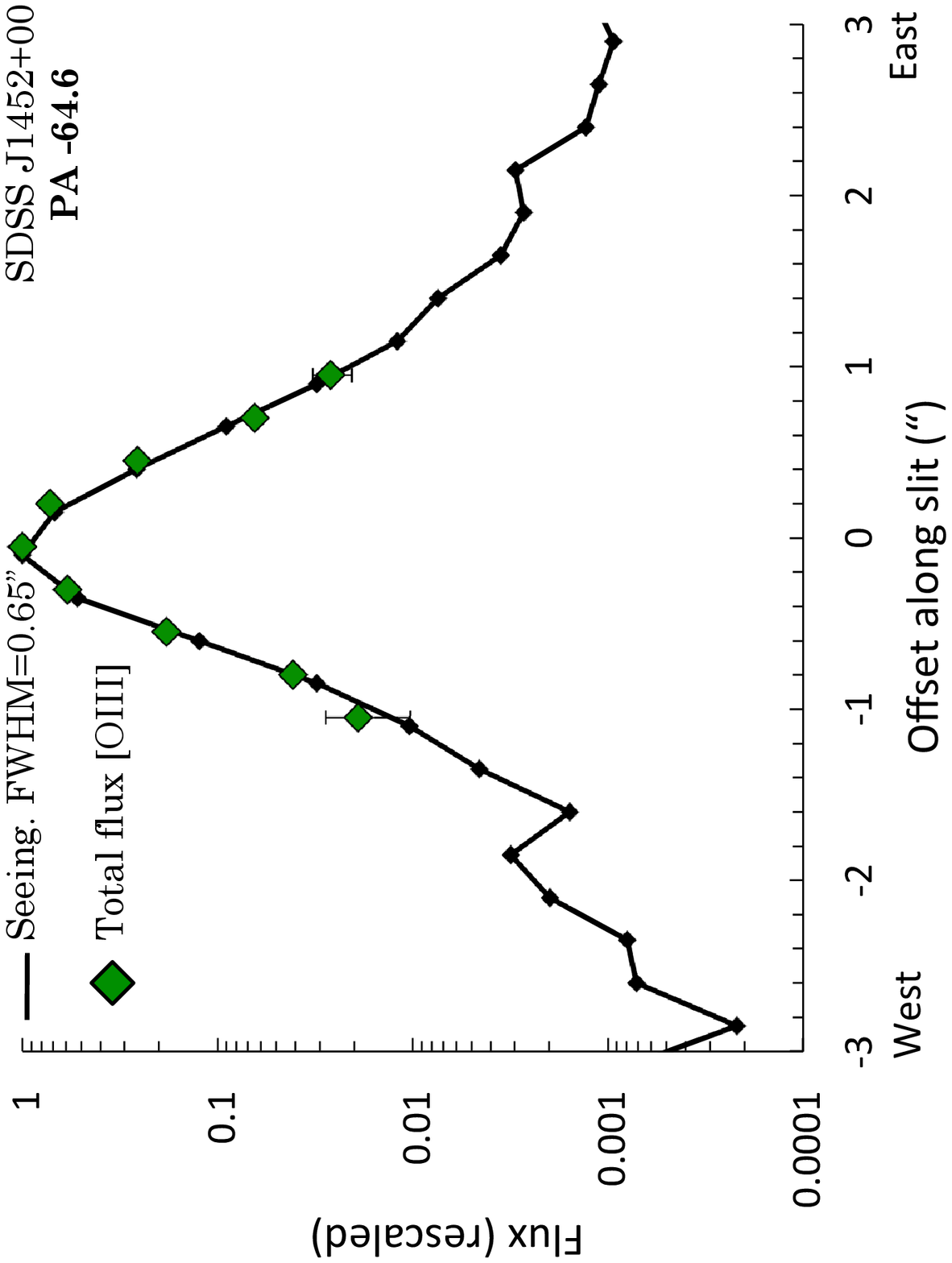}
\vspace{7.5in}
\caption{Spatial profile of the [OIII] $\lambda$5007 emission (also [OII]
  $\lambda$3727 in some cases) along the slit, compared against the spatial profile of the seeing disc.}
\label{spat}
\end{figure*}

Although several strong nebular emission lines are detected in our
spectrum, we find no evidence for  any spatial extension  along our
slit PA. [OII] is not plotted for this object because the strong
adjacent stellar features 
prevent an accurate measurement of the line flux at different spatial
locations (Fig. ~\ref{nuc1336}), but there is no evidence of spatial extension for this line either.  Indeed,
the [OIII] emission  is rather more centrally-peaked than the
5000-6000 \AA~ continuum emission. It is also narrower
(Fig. ~\ref{spat})  than the  seeing size implied by the images and
DIMM values ($\sim$0.6-0.7\arcsec; Table  \,\ref{tab:obs}).

\subsubsection{SDSS J1416-02}

The broad and narrow band images of this HLSy2 show  no clear  tidal
features or possible interacting neighbours
(Fig.~\ref{fig_1416}). However, due to its non-uniform surface
brightness distribution, with an elongation towards the north west, we
consider it plausible that this is a post-interacting system. 

The long slit spectrum  shows  spatially extended
emission lines emitted by an EELR (Fig. ~\ref{fig_1416} and (Fig. ~\ref{spat}), with the strongest line [OIII] $\lambda$5007 having a
total extent of 6\arcsec~ (27 kpc) and maximum extension from the continuum centroid of $\sim$3.25\arcsec or 14.5 kpc.  The spatial peaks of the
emission lines have a small but significant offset of
$\sim$0.3\arcsec~ (1.3 kpc) East of the continuum peak. Across the
observed spatial extent of the nebulosity, the [OIII] $\lambda$5007 /
H$\beta$ flux ratio is within the range 5-10. The detection of the HeII$\lambda$4686 line
on the Western side of the EELR confirms that the  nebula is photoionized by the AGN.

The lines are very narrow in the EELR.  [OIII] is unresolved on the West side of the nebula, with FWHM$\la$200 km s$^{-1}$), while 
FWHM$\la$250 km s$^{-1}$ on the East side, taking slit effects into account.

\subsubsection{SDSS J1452+00}
The broad band image of this HSy2 reveals an arm or tail of faint
emission extending out to $\sim$4\arcsec~ (18 kpc) North West
(Fig.~\ref{fig_1452}).  The association of this extended source is
confirmed by its detection in [OIII] emission at a similar redshift to
the HSy2.  

The [OIII] spatial profile (Fig.~\ref{spat}) is dominated by a compact
unresolved (FWHM$\sim$0.6\arcsec) source. 
In addition, very faint [OIII] extended emission (undetected in the
figure) is detected up to $\sim$3.5\arcsec or 16 kpc from the
continuum centroid (see 2-dim 
spectrum in Fig.~\ref{fig_1452}). This emission
overlaps with the tidal  feature discussed above. The detection of
relatively strong continuum suggests that it is forming
stars. Unfortunately, the spectrum of this region is noisy and the
line FWHM and ratios cannot be  constrained in a useful way. 

\begin{table*}
\centering
\caption{Summary of results. We specify for each object whether an
  EELR has been detected (column 2), and where the observational data
  allow, its nature (column 3). In columns 4 and 5 we give the maximum
  radial extension of the EELR from the AGN, in arcsec and kpc,
  respectively ; in the case of physically distinct emission line
  sources (e.g., star-forming knots), we instead give their offset
  from the AGN. We also specify whether morphological signatures of
  galaxy interactions have been detected in our images (column 6).} 
\begin{tabular}{llllll}
\hline
Galaxy & Feature & Spectrum & Max. Ext. or Dist. & Max. Ext. or Dist. &    Interactions  \\  
            &          &             & (arcsec)   & (kpc)        & \\
(1)&(2)&(3)&(4)&(5)&(6) \\
\hline
SDSS J0903+02 PA2 +63.4        & EELR                              & ? & 4.8 & 22 &    Yes    \\
~~~~~~~~~~~~ PA1 -65.5 &	SF compact object 	&  Composite		&	1.8	&	8 &  \\ \hline
SDSS J0923+01 PA-40.9           &   SF compact object    & SF & 2 & 10  &  Yes \\  \hline
SDSS J0950+01 PA-9.9             & EELR                              & ? & 2.3  & 12  & Yes \\ \hline
SDSS J1014+02  	PA -42.1   &  EELR                             &  ? & 2.8 & 18 &  Yes  \\  \hline
SDSS J1017+03 PA-37.7           &  EELR                             & ? &  3.0   & 17 & Yes \\  
~~~~~~~~~~~~~ 	PA -37.7   &  Compact object          & ? &  3.5 & 20  &  \\  \hline
SDSS J1247+01 PA-60.7           &  EELR                             & ?& 3 & 17 & Yes \\  \hline
SDSS J1336+00 PA-57.4           &  No                              & -- & -- & -- &  No \\  \hline
SDSS J1416-02 PA90                &  EELR
& AGN & 3.25 & 14.5 &  Maybe \\   \hline
SDSS J1452+00 PA64.6             & Tidal tail?                     & ? &  3.5 & 16 & Yes \\ \hline  
\end{tabular}
\label{results}
\end{table*}

\section{Overall Results and Discussion}
We have presented  optical imaging and long slit spectroscopic observations of 9 luminous type
2 AGNs within the redshift range 0.3$<$z$<$0.6. Most (6/9) are high
luminosity  Seyfert 2 (HSy2) and three are type 2 quasars (QSO2). This work expands a similar study by VM11a towards
somewhat less luminous AGN.
Thus far we have described the results on an object-by-object basis. Here, we discuss the overall
results of this study. 

\subsection{Morphological signatures of mergers and interactions}

The  host galaxies of our sample show a variety of optical morphologies, ranging from heavily disturbed 
morphologies suggestive
of a recent or ongoing  major merger, to morphologically regular and unremarkable systems. 

7 out of 9 objects (78\%) show strong morphological evidence for interactions or mergers in the form of disturbed morphologies and/or peculiar features
such as tidal features, amorphous halos, compact emission line knots (see Table ~\ref{results}).  
The two remaining objects -- SDSS J1336-00 (a QSO2) and SDSS J1416-02
(a HLSy2) -- appear as isolated galaxies and have 
no unequivocal signs of mergers/interactions at the depth of our
images. However, SDSS J1416-02 is associated with extended low 
surface brightness continuum emission, with a lopsided flux
distribution that could be interpreted as a post-merger system (among
other possible interpretations). 

In comparison, the detection rate of interactions/mergers found by
VM11a was 5/13 objects (38\%). However, their fraction is a
conservative lower limit (stated by the authors), because they had
only shallow continuum images, and no narrow band emission line
images.  

Indeed, our detection rate of interaction signatures is 
equal to that found by  \cite{bes12} in a complete sample of 20 SDSS selected QSO2 at 3$< z < $0.41 (see also \cite{vm12})
and also similar to that found for  a control sample of quiescent (i.e. non active) early type galaxies of similar mass. \cite{bes12}  found a significant  difference in surface brightness for the interaction features (2 magnitudes brighter for the QSO2). They propose that the mergers witnessed in the comparison sample galaxies could have different progenitors, 
or that the interactions might be viewed at different stages.

\subsection{Extranuclear line emission}
  
Our analysis shows that the spatial distribution of the emission lines
is in general dominated by a compact spatially unresolved
(i.e. consistent with the seeing disk) central source which emits very
strong emission lines associated with the narrow line region. 
In addition, extranuclear line emission of much lower surface brightness
is often detected due to structures of diverse nature: tidal tails,
star forming nuclei/knots/companions and extended
 ionized nebulae. The main results are summarized in Table  \,\ref{results}.
Specifically, we have detected extranuclear line emission  for 8  out of 9
of our sample.   The only exception is SDSS J1336+00, which, on the
other hand, is an isolated galaxy with no obvious signs of
mergers/interactions. The non detection of extranuclear line emission
might be due to a real absence of such emission. However,  unlike for
other objects, given the absence of peculiar/interesting features in the
optical images the slit PA was chosen blindly so that extended lines
along other PA might have been missed.   

Our previous studies  (VM11a and Villar Mart\'\i n et al. 2012)  revealed
extended emission lines associated with 
7/15 QSO2 at similar $z$ 
although this is a gross lower limit, given that the slit was placed
blindly for several objects with no obvious peculiar morphological
features. On the other hand,  we have  shown here that
[OII]$\lambda$3727 
is often more  extended (e.g. SDSS J0950+01, SDSS J1014+02, SDSS
J1247+01) than [OIII]. Our previous results were based on the analysis
of the [OIII] line only. The relatively higher [OII]/[OIII] in the
extended gas suggests that the extranuclear ionized gas has lower
ionization level than the nuclear ionized gas. Moreover, while the gas
in the central, often spatially unresolved region is preferentially
photoionized by the active nucleus, the extranuclear gas is often  at
least partially photoionized by stars (see also McElroy et
al. \citeyear{mce15}). Both facts favor an easier detection of
extranuclear line emission using low ionization lines such as [OII]
compared with [OIII].
 
Integral field spectroscopic works show that extended line emission is
common in QSO2 at this $z$, although these authors studied objects
with very high [OIII] luminosities (Liu et al. \citeyear{liu13a}). 

Extranuclear emission line features other than EELR  (companion
nuclei, knots, tidal tails) are often identified (4/9 in our current
sample or 10/24 in the total sample including our prior
studies). Thus, in $\sim$50\% of the objects emission line
structures whose nature is linked with  mergers/interactions are
detected.  

EELR (which we differentiate from knots, tails, companion nuclei) are
detected for 6/9 objects. The maximum extension from the AGN is in the
range 12-22 kpc, with mean (and median) values of $\sim$17
kpc.   \cite{liu13a} detected extranuclear ionized
nebulae via [OIII]$\lambda$5007  \AA\ emission in most of the QSO2
they studied in a sample of 11 radio quiet QSO2 at similar $z$ to our
sample. They measured radii of the ionized gas nebulae ranging
from $r$=7.5 to $r$=20 kpc, as seen from the 5$\sigma$
detection values. Their mean value is $r=$14 kpc.  
 
Thus, as shown by our previous studies, luminous type 2 AGN are
associated with extranuclear emission line regions, which are often a
complex mixture of tidal/interaction/companion features and EELR, with
a mixture of excitation mechanisms (AGN related processes and stellar 
photoionization) whose relative contribution varies spatially (see
also McElroy et al. \citeyear{mce15}). These regions therefore cannot
be interpreted as a single gaseous structure where the morphology,
kinematics and excitation are determined by a single mechanism. 

\subsection{Ongoing star formation}

VM11a found evidence for recent star formation in the 
  neighbourhood of  5/14 objects. All show signs of
  mergers/interactions. The star formation is happening in general in
  companion galaxies, knots, and/or nuclei.  Definite (i.e.,
  confirmed by the emission line spectrum) evidence for recent star
  formation is confirmed in two 
objects of the sample investigated here: SDSS 0903+02 and SDSS
J0923+01, both with associated compact star forming knots, and
possibly also SDSS 1017+03. All three also show evidence 
of interactions. These fractions are lower limits, given the obviously
limited spatial coverage and the lack of sufficient line ratio
information for diagnostics of the ionizing mechanisms. It is clear
that at least part of the extended SF occurring  in luminous type 2
AGN is triggered by mergers/interactions. A more complete study in two
spatial dimensions would probably reveal  recent star formation at
least in some objects (e.g. disk galaxies) not necessarily related  to
this type of processes   (e.g. McElroy et al. 2015). 
To ascertain the presence of a young stellar population in the nuclear 
regions it would be necessary to fit the optical nuclear spectrum 
with spectral synthesis modeling techniques (e.g. Tadhunter et
al. 2011); Bessiere et al. 2016, in prep.).

\section{Conclusions}

We have presented  optical imaging and long slit spectroscopic
observations of 9 luminous type 
2 AGNs within the redshift range 0.3$<$z$<$0.6 based on VLT-FORS2
data. Six out of the nine objects are  high luminosity  Seyfert 2
(log($\frac{L_{[OIII]}}{L_{\odot}})<8.3$) and three are QSO2 
(log($\frac{L_{[OIII]}}{L_{\odot}})>8.3$). This is an extension of the
work presented in Villar Mart\'\i n et al. (2011a,b), who studied mostly
QSO2, and we have thus enlarged the sample towards somewhat less
luminous type 2 AGN.  

\begin{itemize}

\item {\it Signatures of mergers/interaction}. 7 out of 9 objects
  (78\%) show clear morphological evidence for interactions or mergers
  in the form of disturbed morphologies and/or peculiar features 
such as tidal tails, amorphous halos, compact emission line knots,
etc.   This rate of interaction is consistent with other relevant
studies of (more luminous) QSO2 at similar $z$, suggesting the merger
rate is independent of the AGN luminosity at the high end of the AGN
luminosity function (QSO2 and HLSy2). 

\item {\it Extended line emission}. The emission line spatial profiles
  are dominated by a bright compact, usually spatially unresolved
  central source. 
In addition, much fainter, extranuclear emission line features are
detected associated with 8/9 objects. They are of a diverse nature:
EELR (6/9 objects) of typical radial sizes 12-22 kpc (consistent with
related works focused on more luminous objects) as well as features
related to mergers/interactions such as star forming compact knots
and tidal tails (4/9).  There is a mixture of excitation mechanisms
(AGN related processes and stellar photoionization) whose relative
contribution varies spatially.  While the emission line spectrum of
the ionized gas near the central engine ($R\la$few kpc) is clearly
excited by AGN related processes,  stellar photoionization can also
be present in the extranuclear ionized gaseous structures. Moreover,
we have found evidence, based on the [OII]/[OIII] line flux ratio,
that the extranuclear ionized gas is often in a lower ionization state
than the nuclear ionized gas. In addition, the [OII] emission is often
more spatially extended than [OIII], suggesting that low ionization
lines (such as [OII]) might be relatively more efficient than [OIII]
for detecting extranuclear and extended emissionn line structures in
QSO2. 

\end{itemize}

\section*{Acknowledgments}

We thank the staff at Paranal Observatory for their support during the
observations. AH acknowledges Funda\c{c}\~{a}o para a Ci\^{e}ncia e a
Tecnologia (FCT) support through UID/FIS/04434/2013, and through
project FCOMP-01-0124-FEDER-029170 (Reference FCT
PTDC/FIS-AST/3214/2012) funded by FCT-MEC (PIDDAC) and FEDER
(COMPETE), in addition to FP7 project PIRSES-GA-2013-612701. AH also
acknowledges a Marie Curie Fellowship co-funded by the FP7 and the FCT
(DFRH/WIIA/57/2011) and FP7 / FCT Complementary Support grant
SFRH/BI/52155/2013. MVM and SA acknowledge support from the Spanish Ministerio de
Econom\'\i a y Competitividad through the the grants AYA2012-32295 and
AYA-2012-39408-C02-01. CRA is supported by a Marie Curie Intra
European Fellowship within the 7th European Community Framework
Programme (PIEF-GA-2012-327934). RGD acknowledges support through the
grant AYA2010-15081.

\newcommand{\noopsort}[1]{}

\end{document}